\documentclass[twocolumn,showpacs,preprintnumbers,amsmath,amssymb,superscriptaddress]{revtex4}

\usepackage{graphicx}
\usepackage{dcolumn}
\usepackage{bm}
\usepackage{latexsym}

\begin{document}

\preprint{astro-ph/0209140; NAOJ-Th-Ap 2002,No.27; KUNS-1799
}

\title{Observational Consequences of 
Evolution of Primordial Fluctuations 
in Scalar-Tensor Cosmology}

\author{Ryo Nagata}
\email{nagata@th.nao.ac.jp}
\affiliation{%
Division of Theoretical Astrophysics, National Astronomical Observatory,
2-21-1, Osawa, \\
Mitaka, Tokyo 181-8588, Japan }
\affiliation{
Department of Physics, Kyoto University, Kyoto 606-8502, Japan}
\author{Takeshi Chiba}%
\affiliation{
Department of Physics, Kyoto University, Kyoto 606-8502, Japan}
\author{Naoshi Sugiyama}
\altaffiliation[Also at ]{Max Planck Institute for Astrophysics,
Karl-Schwarzschild-Str. 1 Postfach 1317 D-85741 Garching, Germany}
\affiliation{%
Division of Theoretical Astrophysics, National Astronomical Observatory,
2-21-1, Osawa, \\
Mitaka, Tokyo 181-8588, Japan }

\date{\today}

\begin{abstract}
Evolution of primordial fluctuations in a Brans-Dicke type scalar-tensor
gravity theory is comprehensively investigated.  
The harmonic attractor model, 
in which the scalar field has its 
harmonic effective potential in the Einstein conformal frame and the theory 
relaxes toward Einstein gravity with time, is considered. 
The evolution of adiabatic initial perturbations in flat SCDM models is 
examined from the radiation-dominated epoch up to the present. 
We discuss how the scalar-tensor gravity affects 
the evolution of metric and matter perturbations, mainly focusing on the 
observational consequences, i.e., the matter power spectrum and 
the power spectrum of cosmic microwave background temperature. 
We find that the early time deviation is 
characterized only by the large static gravitational constant 
while the late time behavior is qualitatively different from that in 
Einstein gravity because the time variation of the gravitational
constant and its fluctuation have non-negligible effects. 
The attracting scalar-tensor gravity affects only small scale
modes due to its attracting nature, the degree of which is far beyond
the post-Newtonian deviation at the present epoch. 

\end{abstract}


\maketitle

\section{INTRODUCTION}

The existence of massless scalar partners associated with 
the tensor field of Einstein gravity 
is generically predicted by the recent attempts 
toward unifying all elementary forces in nature 
based on supergravity, superstrings \cite{GSW}, and other higher 
dimensional gravity theories. 
In these theories, time variations of fundamental constants 
such as the gravitational constant and the fine structure constant 
are naturally introduced. 
A recent claim of the time varying fine structure constant 
from observations of QSO absorption lines\cite{Webb} may 
be a piece of evidence.

Scalar-tensor gravity theories, whose original version was 
proposed by Jordan \cite{J} and Brans and Dicke \cite{BD}
and were extended in a more general framework later \cite{BW}, 
naturally provide coupling between the massless scalar fields and the tensor 
field of Einstein gravity.  
Scalar-tensor gravity theories are 
almost only possible alternatives to Einstein gravity.
Moreover, scalar-tensor theories may supply a new approach 
to implementing the inflationary scenario (called extended inflation)
\cite{LS}-\cite{L}. The time-varying gravitational ``constant'' which is
their distinctive feature slows the inflationary expansion from
exponential to power-law in time, and then the inflationary epoch has 
finite period, thereby solving the so-called ``graceful exit'' problem. 
Furthermore, scalar-tensor theories provide a natural framework of 
realizing the time-variation of fundamental constants (gravitational 
constant) via the dynamics of the Brans-Dicke dilaton (for review see 
\cite{CU}). 

In the Jordan-Brans-Dicke theory \cite{J},\cite{BD} (hereafter, we refer
to it as the Brans-Dicke theory for simplicity) which is the simplest 
example of scalar-tensor theories, a constant parameter $\omega$ is 
introduced. In the limit $\omega \rightarrow \infty$, the gravitational 
constant can not change and Einstein gravity is recovered. 
Although scalar-tensor theories including the Brans-Dicke theory are 
compatible with Einstein gravity in several aspects, 
they have many deviations from it. Weak-field experimental 
tests in solar-system have constrained the post-Newtonian deviation 
from Einstein gravity, $\omega > 500$ \cite{REA},\cite{CMW}. 
Measurement of the signal time delay of millisecond pulsars 
or the light deflection of quasars may raise this limit \cite{CMWGQ}. 

In more general scalar-tensor theories, $\omega$ can vary depending 
on the scalar field. In cosmological models based on such theories,
 it has been pointed out that there is generally 
an attractor mechanism that drives
 $\omega$ to $\infty$ in the late cosmological epochs \cite{DN}. 
The nature of gravity can be significantly different in 
the early universe. Hence, information on the different cosmological 
epochs may constrain such theories. A simple and natural extension 
of the Brans-Dicke theory to the attractor model is the harmonic attractor 
model in which the scalar field has a quadratic effective potential of 
a positive curvature in the Einstein conformal frame. 
The analysis of big-bang nucleosynthesis (BBN) in this model \cite{DP} 
restricts two parameters characterizing the potential (its curvature 
$\beta$ and today's gradient $\alpha_0$). 
It is concluded that the BBN limit on the possible deviation from 
Einstein gravity ($2\omega_0+3={\alpha_0}^{-2}$) is much stronger 
than the present observational limits in large $\beta(>0.3)$ models. 
Aside from BBN, we have another source of the information about the 
early universe that is the cosmic microwave background 
(CMB). The trace of primordial fluctuation can be seen clearly in the CMB 
anisotropy spectrum where the information on the early universe 
up to the last scattering time ($z \sim 1000$) is integrated 
on the acoustic peaks. The precise CMB data which will be provided 
in near future \cite{MAP},\cite{PLANCK} should further 
constrain the allowed parameter region. 
Toward probing gravity theories by CMB data, we examine the cosmological 
perturbation evolution in the context of scalar-tensor gravity theories.

Our primary goal is to clarify the growth process of perturbations 
in the scalar-tensor cosmology and understand the influence on the
resultant observable power spectra. We also revisit the 
perturbations of the Brans-Dicke theory which has been investigated by 
several authors \cite{NARI,LMB,CK}.
Although the contemporary plausible 
model is a vacuum dominated model, we adopt flat standard CDM (SCDM) 
models for three reasons. The first reason is that the universe 
necessarily experiences matter domination once at least. 
Therefore, we should begin by revealing the effect of the scalar-tensor 
gravity up to the matter-dominated epoch without confusing curvature or 
vacuum energy effect with it. 
The second reason is that the dependence of the scalar evolution 
on the matter density ($\Omega_0 h^2$) is quite simple as shown in Section IV. 
The final reason is that, if we take account of the current constraint 
for $\omega_0$, the attracting scalar-tensor gravity virtually becomes 
Einstein gravity before vacuum domination and hence any extra process 
originating from the scalar-tensor gravity would not occur after vacuum 
domination. 

This paper is organized as follows: 
In Section II, the field equations and the background cosmological
evolution equations in the scalar-tensor theory are described. 
In Section III, we present the cosmological 
perturbation theory based on the scalar-tensor gravity. 
We demonstrate the numerical calculations in Section IV. 
The resultant matter power spectra and the 
CMB temperature anisotropy spectra are illustrated in Section V. 
Finally, some conclusions are in Section VI. In Appendix, 
the analytic approximate solutions of the background and the perturbation 
equations in the Brans-Dicke theory are discussed. 

\section{MODEL}
In this section, we describe the scalar-tensor theory and 
the (unperturbed) cosmological model based on it. We do not assume a 
specific form of $\omega(\phi)$ as a function of the scalar field 
$\phi$ until performing numerical calculations.

\subsection{Field equation}
The action describing a general massless tensor-monoscalar theory is 
\begin{eqnarray}
S = \frac{1}{16\pi} \int_{}^{} d^4x \hspace{0.2em}
\sqrt[]{-g} \biggl[\hspace{0.1em} \phi R - \frac{\omega(\phi)}{\phi}
(\nabla \phi)^2 \hspace{0.1em} \biggr] \nonumber \hspace{3.0em} \\
+ S_m [\hspace{0.1em} \psi,g_{\mu \nu} \hspace{0.1em}], \label{eq1}
\end{eqnarray}
where $R = g^{\mu \nu} R_{\mu \nu}$ denotes the scalar curvature of 
the metric $g_{\mu \nu}$.
The last term in Eq.(\ref{eq1}) denotes the action of the matter which
is a functional of the matter variable $\psi$  and the metric
$g_{\mu\nu}$. $\omega(\phi)$ is a function of $\phi$ and it 
works as the coupling parameter of the nonminimal coupling. 
We redefine the scalar field so that it is dimensionless
\begin{eqnarray}
\phi \rightarrow \frac{\phi}{G},
\end{eqnarray}
where $G$ is the Newtonian gravitational constant measured today.
The field equations are 
\begin{eqnarray}
R_{\mu\nu} - \frac{1}{2} g_{\mu\nu} R = \frac{8\pi G}{\phi}T_{\mu\nu} 
+ \frac{1}{\phi} ( \nabla_{\mu}\nabla_{\nu}\phi - g_{\mu\nu}\Box\phi) \nonumber \\
+ \frac{\omega}{\phi^2}
\bigl\{ \nabla_{\mu}\phi \nabla_{\nu}\phi - \frac{1}{2} g_{\mu\nu}
(\nabla\phi)^2 \bigr\} 
\label{eq3}
\end{eqnarray}
and
\begin{eqnarray}
R + 2 \frac{\omega}{\phi} \Box \phi &=& - \Bigl(
\frac{1}{\phi} \frac{d\omega}{d\phi} - \frac{\omega}{\phi^2}  \Bigr)
(\nabla\phi)^2, \label{eq4}  \\
\nabla_{\nu} T^{\nu}_{\mu} &=& 0, \label{eq5}
\end{eqnarray}
where $\nabla_{\mu}$ denotes the covariant derivative defined by the  
$g_{\mu\nu}$, $\Box \equiv g^{\mu\nu} \nabla_{\mu} \nabla_{\nu}$,
and $T^{\nu}_{\mu}$ is the matter stress energy tensor defined as follows
\begin{eqnarray}
T^{\mu \nu} & \equiv & \frac{2}{\sqrt[]{-g}}
\frac{\delta S_m [\hspace{0.2em} \psi,g_{\mu \nu} \hspace{0.1em}]}
{\delta g_{\mu \nu}}.
\end{eqnarray}
{The} energy-momentum conservation is satisfied as Eq.(\ref{eq5})
because ordinary matter fields do not directly couple to the scalar field. 
Combining Eqs.(\ref{eq3})and(\ref{eq4}), the equation of motion 
for $\phi$ becomes
\begin{eqnarray}
\Box\phi &=& \frac{1}{2\omega+3} \bigl\{ 8\pi G \hspace{0.1em} T^{\mu}_{\mu} 
- \frac{d \omega}{d \phi} (\nabla\phi)^2 \bigr\}. \label{eq7}
\end{eqnarray}

\subsection{Background cosmological equation}
The unperturbed cosmological spacetime metric is
\begin{eqnarray}
ds^2 = a^2(\eta)( -d\eta^2 + \gamma_{ij} dx^i dx^j),
\end{eqnarray}
where $\gamma_{ij}$ is the metric on the comoving 
homogeneous isotropic 3-space of a constant spatial curvature $K$.
The components of the unperturbed stress energy tensor are
\begin{eqnarray}
T^0_{0}=-\rho, \hspace{1.5em} T^0_{i}=0, \hspace{1.5em} T^i_{j}=p 
\hspace{0.1em} \gamma^i_{j},
\end{eqnarray}
where $\rho$ and $p$ are the total energy density and pressure, respectively.
The background evolution equations are
\begin{eqnarray}
\rho' &=& -3 \frac{a'}{a} (\rho+p) , \label{eq10} \\
\Bigl( \frac{a'}{a} \Bigr)^2 + K &=& \frac{8\pi G \rho a^2}{3\phi} - \frac{a'}{a} \frac{\phi'}{\phi}
+ \frac{\omega}{6} \Bigl( \frac{\phi'}{\phi} \Bigr)^2, \label{eq11} \hspace{4.0em} \\
\phi'' + 2 \frac{a'}{a} \phi' &=& \frac{1}{2\omega+3}
\Bigl\{ 8\pi G a^2(\rho - 3p) - {\phi'}^2 \frac{d \omega}{d \phi} \Bigl\},
 \label{eq12}
\end{eqnarray}
where the prime denotes a derivative with respect to $\eta$.
Although Eq.(\ref{eq10}) is the same as that in Einstein gravity as 
noted previously, the expansion rate $a'/a$ is given by Eq.(\ref{eq11}).
Non-relativistic matter always contributes to the first term 
in the right hand side of Eq.(\ref{eq12}).
The requirement that today's gravitational constant be in agreement with
Newton's constant determines the present value of $\phi$ as
\begin{eqnarray}
\phi_0 = \frac{4+2\omega_0}{3+2\omega_0}, \label{eq13}
\end{eqnarray}
where $\phi_0$ and $\omega_0$ denote the present value of $\phi$ and
 $\omega(\phi)$, respectively.

\section{SCALAR-TENSOR COSMOLOGICAL PERTURBATION}
Following the gauge invariant formulation of the cosmological 
perturbation theory in Einstein gravity \cite{BARD},\cite{KS}, 
we formulate the gauge invariant perturbation theory 
in the scalar-tensor gravity.

\subsection{Perturbed quantities}
Perturbations of scalar quantities can be
 expanded by the harmonic functions Y defined as
\begin{eqnarray}
(\Delta+k^2)Y &=& 0,
\end{eqnarray}
where $-k^2$ represent their eigen values of Laplace-Beltrami operator
$\Delta$ on the comoving homogeneous 3-space. Scalar type components in 
perturbations of vector quantities are expanded by
\begin{eqnarray}
Y_{i} & \equiv & -k^{-1} Y_{ | i},
\end{eqnarray}
where ${_{ | i}}$ denotes a covariant derivative associated with $\gamma_{ij}$.
Those in perturbations of tensor quantities are expanded by 
\begin{eqnarray}
Y \gamma_{ij}
\end{eqnarray}
and
\begin{eqnarray}
Y_{ij}  \equiv  \bigl( k^{-2}Y_{ | ij} + \frac{1}{3} \gamma_{ij}Y \bigr),
\end{eqnarray}
which correspond to the trace and the traceless part, respectively.
In this paper,
we will omit the summation symbols and the eigenvalue indices because 
there is no coupling among the different $k$-modes.
{Denoting} the perturbed metric by $\tilde{g}_{\mu\nu}$,
the components of the perturbed metric are written as
\begin{eqnarray}
\tilde{ g }_{00} &=& -a^2 (1+2AY), \\
\tilde{ g }_{0j} &=& -a^2 B Y_{j}, \\
\tilde{ g }_{ij} &=&  a^2 \bigl\{ (1+2H_{L}Y) \gamma_{ij} + 2
H_{T}Y_{ij} \bigr\}.
\end{eqnarray}
The perturbed scalar field is
\begin{eqnarray}
\tilde{ \phi } = \phi + \chi Y.
\end{eqnarray}
The perturbed energy momentum tensor $\tilde{T}^{\mu}_{\nu}$
has the following components:
\begin{eqnarray}
\tilde{ T }^{0}_{0} &=& -\rho (1+\delta Y), \\
\tilde{ T }^{0}_{j} &=& (\rho+p)(v-B)Y_{j}, \\
\tilde{ T }^{j}_{0} &=& -(\rho+p)vY^{j}, \\
\tilde{ T }^{i}_{j} &=&  p \bigl\{ (1+\pi_{L}Y) \gamma^{i}_{j} +
\pi_{T}Y^{i}_{j} \bigr\},
\end{eqnarray}
where $\delta$, $v$, $\pi_L$ and $\pi_T$ are the perturbations of the 
energy density, the spatial velocity and the isotropic and the
anisotropic pressure, respectively. All of the above perturbation 
quantities are functions only of time. When we consider the background
and the perturbed universe, there is gauge freedom to fix the 
correspondence between the spacetime points in the 
two spacetimes. Denoting a scalar
 type infinitesimal gauge transformation by the vector
\begin{eqnarray}
\xi^{\mu} &=& ( \hspace{0.5em} TY,\hspace{0.5em} LY^j\hspace{0.5em}),
\end{eqnarray}
where $T$ and $L$ are arbitrary functions of time which has the same
order as perturbation variables, we obtain the changes of perturbations
under the gauge transformation as
\begin{eqnarray}
\bar{A} &=& A-T'-(a'/a)T, \\
\bar{B} &=& B+L'+kT, \\
\bar{H}_L &=& H_L - (k/3)L - (a'/a)T, \\
\bar{H}_T &=& H_T + kL,
\end{eqnarray}
\begin{eqnarray}
\bar{\delta} &=& \delta - (\rho'/\rho)T, \\
\bar{v} &=& v+L', \\
\bar{\pi}_L &=& \pi_L - (p'/p)T, \\
\bar{\pi}_T &=& \pi_T,
\end{eqnarray}
and
\begin{eqnarray}
\bar{\phi} &=& \chi - \phi'T,
\end{eqnarray}
where the variables with bar represent the gauge transformed values.
To treat the evolution of perturbations without unphysical gauge modes, 
we deal only with gauge-invariant variables. We can construct 
gauge-invariant variables by combining perturbation variables. 
For one perturbation variable, there are several gauge invariants each 
of which corresponds to the perturbation in a specific gauge. 
Here, we introduce two gauge invariants for each of the metric, the
scalar field and the density perturbations. 
The one denoted by subscript $s$ corresponds to the perturbation in the
Newtonian shear free gauge (where $H_T=B=0$). 
The other denoted by subscript $c$ corresponds to the perturbation in
the total matter comoving gauge (where $B=v$). 
Let us proceed to the definition. We can employ two
independent variables $\Psi$ and $\Phi$ for metric perturbations. 
The gauge invariant perturbation variables associated with the (0-0)
component of the metric are defined as
\begin{eqnarray}
\Psi_s  \equiv  A + k^{-1}(a'/a)(B - k^{-1} {H_{T}}') \hspace{4.2em} \nonumber \\
+ k^{-1}(B' - k^{-1} {H_{T}}''),
\end{eqnarray}
\begin{eqnarray}
\Psi_c  \equiv  A + k^{-1}(a'/a)(B - v) + k^{-1}(B' - v').
\end{eqnarray}
These correspond to the perturbations of the gravitational potential in
the two gauges. On the other hand, the gauge invariants associated
{with} the spatial component of the metric are
\begin{eqnarray}
\Phi_s & \equiv & H_{L} + \frac{1}{3} H_{T} + k^{-1}(a'/a)(B -
k^{-1}{H_{T}}'),\\
\Phi_c & \equiv & H_{L} + \frac{1}{3} H_{T} + k^{-1}(a'/a)(B - v),
\end{eqnarray}
which represent the perturbations of the intrinsic spatial curvature in
the two gauges. Similarly we introduce two gauge invariants of 
the scalar field perturbation
\begin{eqnarray}
X_{s}    & \equiv & \chi - k^{-1}  \phi' (k^{-1}{H_{T}}' - B), \\
X_{c} & \equiv & \chi - k^{-1}  \phi' (v - B).
\end{eqnarray}
The gauge invariants of the energy momentum tensor perturbations
are defined as follows:
\begin{eqnarray}
\Delta_{s} & \equiv & \delta + 3(1+w)(a'/a)k^{-1}(k^{-1}{H_{T}}' - B),\\
\Delta_{c} & \equiv & \delta + 3(1+w)(a'/a)k^{-1}(v-B), 
\end{eqnarray}
\begin{eqnarray}
V & \equiv & v - k^{-1}{H_{T}}',
\end{eqnarray}
\begin{eqnarray}
\Gamma & \equiv & \pi_{L} - \frac{ c_{s}^2 }{w} \delta,
\end{eqnarray}
\begin{eqnarray}
\Pi & \equiv & \pi_{T},
\end{eqnarray}
where
\begin{eqnarray}
w &=& p/\rho, \\
c_{s}^2 &=& p'/\rho'.
\end{eqnarray}
The conventional photon temperature fluctuation variable is
\begin{eqnarray}
\Theta_0 &=& \frac{1}{4} \Delta_{s\gamma},
\end{eqnarray}
where the subscript $\gamma$ means the photon component.

\subsection{Perturbation equation}
The perturbation equations of Eq.(\ref{eq3}) are
\begin{widetext}
\begin{eqnarray}
\frac{2}{a^2} \Biggl[ 3 { \Bigl( \frac{a'}{a} \Bigr) }^2 \Psi_s - 3 \frac{a'}{a} {\Phi_s}'
- (k^2 - 3K)\Phi_s \Biggr] \hspace{23.5em} \nonumber \\
= - \frac{8\pi G}{\phi} \rho\Delta_{s}
+ \frac{3}{a^2\phi} X_{s}\Bigl\{ { \Bigl( \frac{a'}{a} \Bigr) }^2 + K \Bigr\}
- \frac{1}{a^2\phi} \Biggl[ \Bigl\{ 6\Bigl( \frac{a'}{a} \Bigr)\Psi_s - 3\Phi_s' \Bigr\}\phi' 
- 3\Bigl( \frac{a'}{a} \Bigr)X_{s}'-k^2X_{s} \Biggr] \hspace{0.5em} \nonumber \\
- \frac{1}{2a^2} X_{s} { \Bigl( \frac{\phi'}{\phi} \Bigr) }^2 \frac{d \omega}{d \phi}
+ \frac{\omega}{a^2\phi}
\Biggl[
\frac{1}{2} \Bigl( \frac{\phi'}{\phi} \Bigr)^2 X_{s}
 - \Bigl( \frac{\phi'}{\phi} \Bigr) X_{s}' + \frac{ {\phi'}^2 }{\phi} \Psi_s 
\Biggr], \label{eq49}
\end{eqnarray}
\begin{eqnarray}
\frac{2}{a^2} \Biggl[ k \frac{a'}{a}  \Psi_s
- k {\Phi_s }'
  \Biggr]
= \frac{8\pi G}{\phi} (\rho+p)V
- \frac{k}{a^2\phi} \Bigl\{ \Bigl( \frac{a'}{a} \Bigr) X_{s} + \phi' \Psi_s - X_{s}' \Bigr\}
+ \frac{k \omega}{a^2\phi} \Bigl( \frac{\phi'}{\phi} \Bigr)X_{s}, \hspace{5.2em}
 \label{eq50}
\end{eqnarray}
\begin{eqnarray}
\frac{2}{a^2} \Biggl[ \frac{a'}{a} {\Psi_s}' + 
\Bigl\{ 2 { \Bigl( \frac{a'}{a} \Bigr) }'
+ { \Bigl( \frac{a'}{a} \Bigr) }^2 - \frac{k^2}{3}  \Bigr\} \Psi_s - \Phi_s''
- 2 \frac{a'}{a} \Phi_s' - \frac{k^2}{3} \Phi_s +K \Phi_s
\Biggr]  \hspace{10.0em} \nonumber \\
= \frac{8\pi G}{\phi} (p \Gamma+ \rho c_{s}^2 \Delta_{s}) + \frac{1}{a^2\phi} X_{s}
\Bigl\{ 2{ \Bigl( \frac{a'}{a} \Bigr) }' +{ \Bigl( \frac{a'}{a} \Bigr) }^2 + K \Bigr\}
\hspace{15.0em} \nonumber \\
- \frac{1}{a^2\phi} \Biggl[ 2 \phi'' \Psi_s + \phi' \Bigl\{ \Psi_s' + 2 \frac{a'}{a} \Psi_s
- 2 \Phi_s' \Bigr\}
- X_{s}'' - \frac{a'}{a} X_{s}' - \frac{2k^2}{3} X_{s} \Biggr] \hspace{9.0em} \nonumber \\
+ \frac{1}{2a^2} X_{s} { \Bigl( \frac{\phi'}{\phi} \Bigr) }^2 \frac{d \omega}{d \phi}
- \frac{\omega}{a^2\phi}
\Biggl[
\frac{1}{2} \Bigl( \frac{\phi'}{\phi} \Bigr)^2 X_{s}
 - \Bigl( \frac{\phi'}{\phi} \Bigr) X_{s}' + \frac{ {\phi'}^2 }{\phi} \Psi_s 
\Biggr],
\end{eqnarray}
\begin{eqnarray}
- \frac{k^2}{a^2} \Bigl\{ \Phi_s + \Psi_s \Bigr\}
= \frac{8\pi G}{\phi} p \Pi + \frac{k^2}{a^2\phi} X_{s}. \hspace{26.0em}
\label{eq52}
\end{eqnarray}
\end{widetext}
The perturbation equation of Eq.(\ref{eq7}) is 
\begin{widetext}
\begin{eqnarray}
X_s'' + 2\frac{a'}{a}X_s' + k^2X_s - 2\phi''\Psi_s
- \phi'(\Psi_s'+4\frac{a'}{a}\Psi_s-3\Phi_s') \hspace{24.5em} \nonumber \\
=\frac{a^2}{2\omega+3} \Bigl[ 8\pi G \rho \bigl\{ (1-3c_{s}^2)\Delta_{s} - 3w\Gamma \bigr\}
-\frac{d^2 \omega}{d \phi^2}
X_s a^{-2} {\phi'}^2 - 2\frac{d \omega}{d \phi}  a^{-2}
(\phi' X_s' - {\phi'}^2 \Psi_s) 
- 2\frac{d \omega}{d \phi}  a^{-2} X_s(\phi''+2\frac{a'}{a}\phi') \Bigr].
 \label{eq53}
\end{eqnarray}
\end{widetext}
The perturbation equations of Eq.(\ref{eq5}) are
\begin{eqnarray}
\Delta_c' - 3 \Bigl( \frac{a'}{a} \Bigr) w \Delta_c 
= -3(1+w) \Phi_c' -(1+w) k V \nonumber \\
- 3 \Bigl( \frac{a'}{a} \Bigr) w \Gamma -3 \Bigl( \frac{a'}{a} \Bigr) c_{s}^2 \Delta_c,
\end{eqnarray}
\begin{eqnarray}
V' + \Bigl( \frac{a'}{a} \Bigr) V 
= k\Psi_s \hspace{13.0em} \nonumber \\
+\frac{k}{ 1+w } \Bigl\{ c_{s}^2 \Delta_{c} + w \Gamma 
- \frac{2}{3} \Bigl( 1-\frac{3K}{k^2} \Bigr) w \Pi \Bigr\}. \label{eq55} 
\end{eqnarray}
These are the same as those in Einstein gravity. 
Of course, the Boltzmann equations for photons and neutrinos are also not 
modified. Hence we omit them here. From Eq.(\ref{eq55}), the relation 
between the potential perturbation and pressure perturbations becomes
\begin{eqnarray}
\Psi_c &=& - \frac{1}{ 1+w } \Bigl\{ c_{s}^2 \Delta_{c} + w \Gamma
- \frac{2}{3} \Bigl( 1-\frac{3K}{k^2} \Bigr) w \Pi \Bigr\}.
\end{eqnarray}
Employing the total matter gauge variables and using
\begin{eqnarray}
\frac{a'}{a}  \Psi_c - {\Phi_c }'
= K \frac{V}{k} - \frac{1}{2\phi} \Bigl\{ \Bigl( \frac{a'}{a} \Bigr) X_{c} + \phi' \Psi_c - X_{c}' \Bigr\}
\nonumber \\
+ \frac{\omega}{2\phi} \Bigl( \frac{\phi'}{\phi} \Bigr)X_{c}, \label{eq56}
\end{eqnarray}
we obtain the alternative expression of Eq.(\ref{eq53}) that is
\begin{widetext}
\begin{eqnarray}
X_c'' + 2\frac{a'}{a}X_c' + k^2X_c - 2\phi''\Psi_c
- \phi'(\Psi_c' + \frac{a'}{a}\Psi_c )  
+ k \phi' \Bigl( 1-\frac{3K}{k^2} \Bigr)V + \frac{3}{2} \frac{\phi'}{\phi}
\Bigl\{ \Bigl( \frac{a'}{a} \Bigr)
X_{c} + \phi' \Psi_c - X_{c}' \Bigr\} - \frac{3\omega}{2}
\Bigl( \frac{\phi'}{\phi} \Bigr)^2 X_{c} \nonumber \\
=\frac{a^2}{2\omega+3} \Bigl[ 8\pi G \rho \bigl\{ (1-3c_{s}^2)\Delta_c - 3w\Gamma \bigr\}
-\frac{d^2 \omega}{d \phi^2}
X_c a^{-2} {\phi'}^2 
- 2\frac{d \omega}{d \phi}  a^{-2} (\phi' X_c' - {\phi'}^2 \Psi_c)
- 2\frac{d \omega}{d \phi}  a^{-2} X_c(\phi''+2\frac{a'}{a}\phi') \Bigr].
\label{eq57}
\end{eqnarray}
\end{widetext}
Combining Eqs.(\ref{eq49}), (\ref{eq50}) and (\ref{eq52}), we obtain the generalized Poisson equation in the scalar-tensor
theory as follows:
\begin{widetext}
\begin{eqnarray}
- (k^2 - 3K) \Psi_s 
= \frac{4\pi G \rho \hspace{0.1em} a^2}{\phi}  \Bigl( \Delta_{c} - \frac{2}{\phi} X_c \Bigr)
+\frac{1}{4} \Bigl( \frac{\phi'}{\phi} \Bigr)^2 \frac{d \omega}{d \phi} X_c
+\frac{2\omega+3}{4 \phi} \Bigl( \frac{\phi'}{\phi} \Bigr)
\Bigl\{ X_c'+ 3 \Bigl( \frac{a'}{a} \Bigr) X_{c} - \phi' \Psi_c \Bigr\} 
\hspace{3.0em} \nonumber \\
+(k^2-3K) \Bigl\{ \frac{8\pi G}{\phi} { \Bigl( \frac{a}{k} \Bigr) }^2 p \Pi + \frac{X_c}{2\phi}
+\frac{1}{2k} \frac{\phi'}{\phi} V \Bigr\}.
\label{eq58}
\end{eqnarray}
\end{widetext}
The gravitational redshift factor in the scalar-tensor theory is written as:
\begin{eqnarray}
\Phi_s - \Psi_s = \Upsilon_{\psi} + \Upsilon_{\phi},
\end{eqnarray}
where
\begin{widetext}
\begin{eqnarray}
\Upsilon_{\psi} & \equiv & \frac{8\pi G}{\phi} 
{ \Bigl( \frac{a}{k} \Bigr)}^2 
\Bigl\{ \Bigl( 1-\frac{3K}{k^2} \Bigr)^{-1} \rho \Delta_{c} 
+ p \Pi \Bigr\}, \\
\Upsilon_{\phi \hspace{0.6em}} & \equiv & ( k^2 - 3K )^{-1}
\Biggl[ 
-\frac{16 \pi G \rho \hspace{0.1em} a^2}{\phi} \frac{X_c}{\phi}
+\frac{1}{2} \Bigl( \frac{\phi'}{
\phi} \Bigr)^2 \frac{d \omega}{d \phi} X_c
+ \frac{2\omega+3}{2 \phi} \Bigl( \frac{\phi'}{\phi} \Bigr)
\Bigl\{ X_c'+ 3 \Bigl( \frac{a'}{a} \Bigr) X_{c} - \phi' \Psi_c \Bigr\}
\Biggr]. \hspace{2.2em}
\end{eqnarray}
\end{widetext}
$\Upsilon_{\psi}$ and $\Upsilon_{\phi}$ are the contribution of 
matter perturbations and the scalar field perturbation, respectively. 
If the background spacetime and the initial conditions for
perturbations are given, we can solve the above equations.
The initial conditions for perturbations (solutions on a
 superhorizon scale in the radiation-dominated epoch) are discussed 
in the next section. 

\section{NUMERICAL CALCULATION}
Let us now numerically solve the evolution of the background universe
 and perturbations. We apply our formulation to
 flat SCDM models. The matter components are baryons, cold
dark matters, photons, and three species of massless neutrinos.
(In cold dark matter dominated models, massive neutrinos
 do not significantly change the evolution of $\phi$.) 
In our numerical calculations, the functional form of $\omega(\phi)$ is 
\begin{eqnarray}
2\omega+3 &=& \bigl\{ \hspace{0.1em} {\alpha_0}^2- \beta \ln(\phi/\phi_{0})
\hspace{0.1em}  \bigr\}^{-1}. \label{eq59}
\end{eqnarray}
This $\omega(\phi)$ corresponds to the harmonic effective potential 
of the scalar field in the Einstein conformal frame. 
$\alpha_0$ and $\beta$ in Eq.(\ref{eq59}) are today's potential gradient 
and curvature, respectively \cite{DN},\cite{DP}. If $\beta = 0$, 
this model is reduced to the Brans-Dicke theory. Moreover, the model 
with $\alpha_0 \rightarrow 0, \beta=0$ is Einstein gravity.
To realize the cosmological attractor mechanism, we treat 
the parameter region $\beta \geq 0$ in this paper. 
Although some parameter region is already excluded by solar-system 
experiments and the BBN analysis, we often adopt unrealistically large 
values of $\alpha_0$ and $\beta$ to see the qualitative parameter 
dependence of the results.

\subsection{Boundary condition}
First, we describe the boundary conditions for background variables. 
We employ the conventional definitions of cosmological 
parameters
\begin{eqnarray}
\Omega_0  \equiv  \rho_0 / \rho_c, \hspace{0.8em} 
\Omega_K  \equiv  - K / a_0^2 H_0^2, \hspace{0.8em}
\rho_c  \equiv  3 H_0^2 / 8 \pi G.
\end{eqnarray}
Unless otherwise noted, we fix today's baryon density 
($\Omega_b ( \equiv \rho_{b0} / \rho_c )$) and Hubble parameter ($h$) to $0.03$ and $0.7$, respectively. 
(In this paper, we do not discuss the light element abundance synthesized 
via BBN in the scalar-tensor cosmology. The helium mass fraction is fixed at $23\%$ in our calculations.) 
Then, the density of CDM is determined by $\Omega_{c0} = \Omega_0 - \Omega_{b0} $. 
In the scalar-tensor theory, the relation between the 
cosmological parameters is as follows
\begin{eqnarray}
\frac{\Omega_0}{\phi_0} + \Omega_K - S + \frac{\omega(\phi_0)}{6} S^2  = 1 ,
\end{eqnarray}
where $S \equiv \dot{\phi_0} / \phi_0 H_0$ and 
$\dot{\phi}=\frac{1}{a} \frac{d}{d \eta} \phi$.  
We set $\Omega_K$ in the above equation to be zero since we treat 
only flat models.
As mentioned in Section II, $\phi_0$ must satisfy Eq.(\ref{eq13}).
In the scalar-tensor theory, $\Omega_0 $ is not necessarily unity even
in flat SCDM models. There exists the degree of freedom to determine 
the amplitude of the vacuum solution mode of $\phi'$ ($\propto a^{-2}
(2\omega+3)^{-\frac{1}{2}}$) at the present time. 
As in the precedent study \cite{CK}, we assume that such a mode is
negligible ever since the initial time of our calculations. Then, from 
the particular solution during the matter dominated epoch \cite{DN}, 
the relation that $\Omega_0$ and $S$ must satisfy in 
$ 0 \leq \beta < 3/8$ models becomes 
\begin{eqnarray}
\Omega_0 = \frac{ (1+\alpha_0^2) (1-\partial \varphi_0^2/3) }{(1+\alpha_0\partial\varphi_0)^2} 
 , \hspace{1em} S = - \frac{ 2 \alpha_0 \partial \varphi_0}{(1+\alpha_0\partial\varphi_0)}, \label{eq62}
\end{eqnarray}
where $\partial \varphi_0 = - 3 \alpha_0 (1-r)/ (4\beta)$ and
$r=(1-8\beta/3)^{1/2}$. 
In this paper, we set the initial time $z \sim 10^{8}$ so that the modes
observed today as large scale structures are still before their horizon
entry. Moreover, the energy of electrons and positrons already have no
effect on $\phi$ evolution at that time. We search for the numerical 
solution of $\phi$ satisfying the above boundary conditions by 
iterating the initial value of $\phi, \phi_{ini}$. 
The initial value of $\phi'$ to a given $\phi_{ini}$ which is 
composed of the particular solution is 
\begin{eqnarray}
\phi'_{ini} = \frac{3\phi_{ini}}{4\omega_{ini}+6} \frac{a}{a_{eq}}
\Bigl( \frac{8 \pi G \rho a^2}{3 \phi_{ini}} \Bigr)^{\frac{1}{2}}.
\end{eqnarray}
Here we assume that $\phi$ (and $\omega(\phi)$) are almost constant and
the $\phi'$ terms in Eq.(\ref{eq11}) are negligible at the initial time. 
In fact, the factor $a/a_{eq}$ extremely damps its amplitude and this 
is consistent with these assumptions. 

Next, the initial conditions for perturbation variables should be
noted. Here, the initial condition means the solution on a 
superhorizon scale in the radiation-dominated epoch. 
To obtain the initial solution of $X_c$, we employ the similar procedure
as applied to $\phi'$. In addition, for matter perturbations, we adopt
the same adiabatic initial conditions as those in Einstein gravity. 
Then, the homogeneous solution of Eq.(\ref{eq57}) consists of a constant 
and a decaying mode, whose contribution to the gravitational potential 
perturbation decreases with time. Neglecting them and using the 
initial solutions for matter perturbations, we obtain the growing 
{particular} solution of $X_c$ that is 
\begin{eqnarray}
X_c  = \frac{3\phi_{ini}}{32\omega_{ini}+48} \frac{a}{a_{eq}} 
\frac{5+3f_{\nu}}{7+5f_{\nu}} \Delta_c,
\end{eqnarray}
where $f_{\nu} = \rho_{\nu} / \rho_{\gamma}$. 
This solution contains neutrino anisotropic stress contribution. 
Consistently, this initial solution of $X_c$ does not modify 
the initial solutions for matter perturbations because the contribution
of $X_c$ (and $\phi'$) to metric perturbations is at most comparable to
that of CDM perturbation $O(a/a_{eq})$. 

\subsection{Background}
Here we compare the time evolution of background variables in some
typical models. 

\begin{figure}
\includegraphics[width=\hsize]{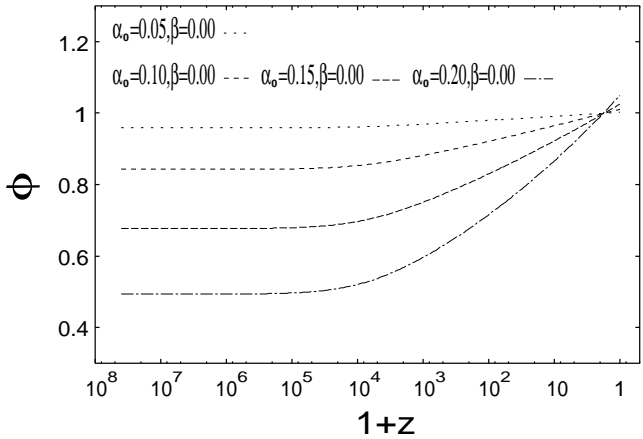}
\caption{The time evolution of $\phi$ in the Brans-Dicke models with 
SCDM parameters where $\Omega_0$ is determined by Eq.(\ref{eq62}). 
The discrepancy at the present is due to the different $\omega$. }
\label{fig:alp_evo}
\end{figure}

\begin{figure}
\includegraphics[width=\hsize]{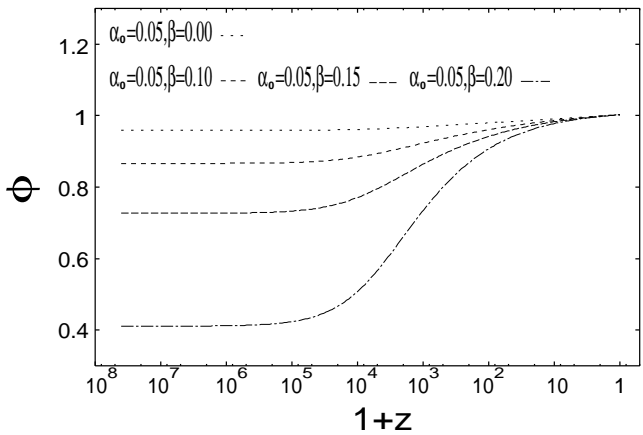}
\caption{The time evolution of $\phi$ in the attractor models with 
SCDM parameters where $\Omega_0$ is determined by Eq.(\ref{eq62}). 
Note that the variation of $\omega$ makes an inflection point around the
equality time (cf. Fig.\ref{fig:hub_evo}). 
The alignment at the present time is due to the equality of $\omega(\phi_0)$. }
\label{fig:phi_evo}
\end{figure}

\begin{figure}
\includegraphics[width=\hsize]{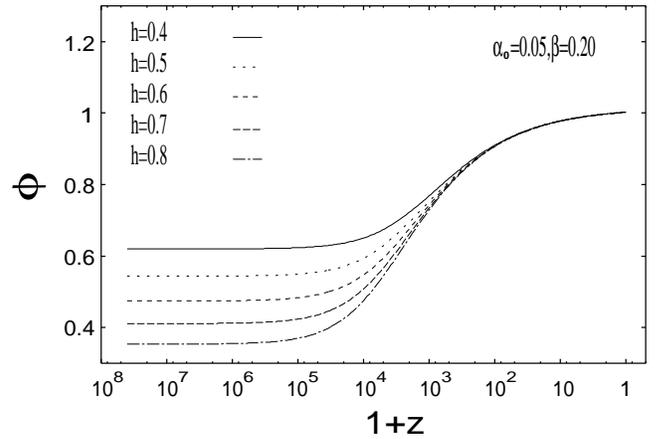}
\caption{The density dependence of $\phi$ history in the attractor
  models with SCDM parameters. $\Omega_0$ is determined by
  Eq.(\ref{eq62}) and $\Omega_b h^2$ is fixed to 0.0147. }
\label{fig:hdep}
\end{figure}

Fig.\ref{fig:alp_evo} shows the $\alpha_0$ dependence of $\phi$ evolution 
and Fig.\ref{fig:phi_evo} shows the $\beta$ dependence of it. 
$\phi$ is frozen during the radiation-dominated epoch and 
begins to roll down at the matter-radiation equality time 
to realize the Newtonian gravitational constant at the present. 
As increasing $\alpha_0$ or $\beta$, we obtain smaller initial $\phi$. 
In the Brans-Dicke (constant $\omega$) models, the late time slope is
monotonic. On the other hand, in the attractor (variable $\omega$)
models, the time variation of $\omega(\phi)$ which regulates 
the amplitude of $\phi'$ (proportional to $(2\omega+3)^{-1}$) 
causes a steep slope immediately after the equality time and a gentle
slope near the present. For reference, the $\Omega_0 h^2$ dependence 
of $\phi$ evolution is illustrated in Fig.\ref{fig:hdep}. 
The shift of equality time simply decreases the initial value of $\phi$
without altering the late time evolution. 
Now, even if we employ a vacuum dominated model, 
the period from vacuum domination is quite short in units of $\ln a$ 
and hence the only influence on $\phi$ is to decrease in today's matter 
density. Thus, in a vacuum dominated model, the history of $\phi$ is
modified only before the equality time 
as the low density SCDM models in Fig.\ref{fig:hdep}. 

\begin{figure}
\includegraphics[width=\hsize]{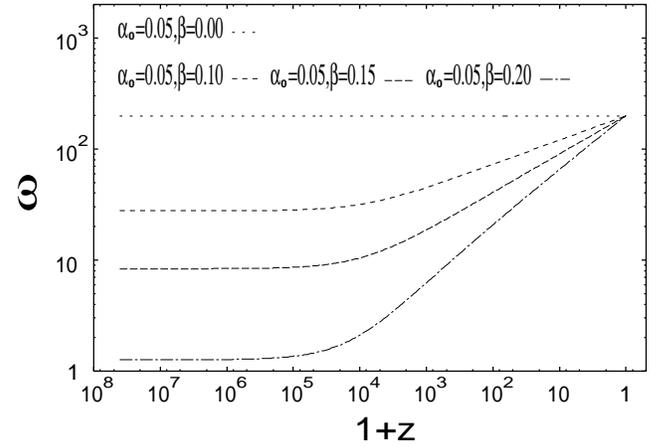}
\caption{The time evolution of $\omega(\phi)$ in the attractor models
  with SCDM parameters. }
\label{fig:om_evo}
\end{figure}

Fig.\ref{fig:om_evo} compares the time evolution of $\omega(\phi)$ in
the attractor models. $\omega$ can vary only after the equality time
because, even in variable $\omega$ models, the variation of $\phi$ is
suppressed before the equality time. The late time power-law behavior 
is well described by
\begin{eqnarray}
2\omega+3 \simeq (2\omega_0+3) \Bigl( \frac{a}{a_0} \Bigr)^{\frac{3}{2}
  (1-r)}.
\end{eqnarray}
$r$ is defined in the previous section. 

\begin{figure}
\includegraphics[width=\hsize]{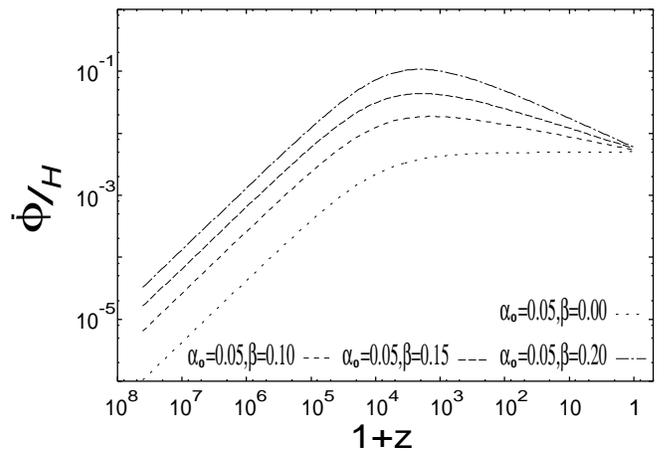}
\caption{The time evolution of $\dot{\phi}/H$ in the attractor models 
with SCDM parameters, where $\dot{\phi}=\frac{1}{a} \frac{d}{d \eta} \phi$. 
The maximums correspond to the inflection points in Fig.\ref{fig:phi_evo}. }
\label{fig:hub_evo}
\end{figure}

Fig.\ref{fig:hub_evo} illustrates the time evolution of $\phi' / (a'/a)$
in the attractor models which also represents the variation of $\phi$ 
during a logarithmic interval of $1+z$. (It should be noted about
$\alpha_0$ dependence that the whole amplitude of these curves is nearly
proportional to ${\alpha_0}^2$.) 
The variability of $\phi$ is summarized as follows:
\begin{eqnarray}
\phi'  \Bigl( \frac{a'}{a} \Bigr)^{-1} = \frac{d \phi}{d \ln a} \sim \left\{ 
\begin{array}{ll}
\frac{1}{\omega}  \frac{a}{a_{eq}} & {\rm RD} \\
\frac{1}{\omega}. & {\rm MD} \\
\end{array}
\right.
\end{eqnarray}
The initial slope represents the ratio of matter energy density to 
radiation energy density. Then, the motion of $\phi$ is negligible and 
does not affect the spacetime metric through either of the minimal 
and the non-minimal coupling. Hence, the scalar-tensor gravity
effectively behaves like Einstein gravity and 
the only difference is the magnitude of the gravitational coupling
"constant". The larger gravitational constant results in the larger
expansion rate and therefore the smaller horizon length. 
On the other hand, in the matter-dominated epoch, $\phi$ can move in the
absence of the suppression factor $O(a/a_{eq})$. 
As a result, the time variation of $\phi$ changes the nature of gravity
qualitatively. For example, $a$ is not proportional to $\eta^2$. 
Especially in the attractor models, such effects appear only in the 
intermediate epochs between the equality and the present 
because the deviation from Einstein gravity is in proportion to
$\omega(\phi)^{-1}$. Thus, the effect of scalar field dynamics is most 
significant around the equality time. 


\subsection{Perturbation}
In this section, the evolution of perturbation variables is shown. 
When comparing the perturbation spectra, we adopt scale invariant 
models without discussing the generation mechanism of perturbations. 
The amplitude of the curvature perturbation ($\Phi_s$) in each model is
initially indistinguishable. Hence, a comparison between models shows 
directly the difference in their evolution. 

\subsubsection{Scalar field perturbation}
\begin{figure}
\includegraphics[width=\hsize]{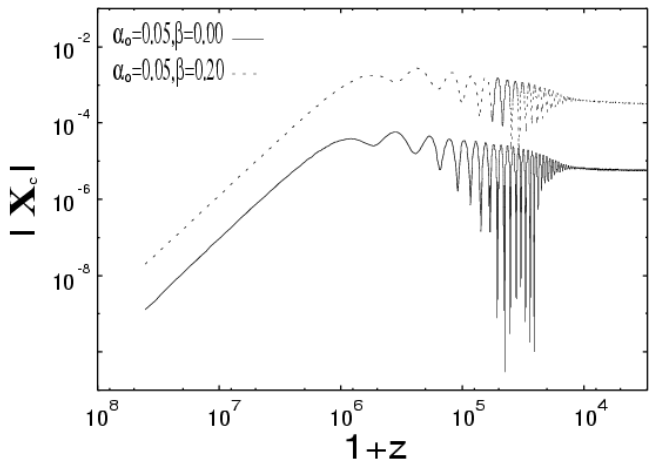}
\caption{The early evolution of $X_c$ on a scale $k=10$ Mpc${}^{-1}$ 
in the scalar-tensor models with SCDM parameters. 
Normalization is arbitrary. Their amplitude is proportional to 
$\omega^{-1}$ (or ${\omega_{ini}}^{-1}$). The discrepancy of their 
horizon entry times is due to the difference in $\phi_{initial}$.} 
\label{fig:xxs}
\end{figure}

\begin{figure}
\includegraphics[width=\hsize]{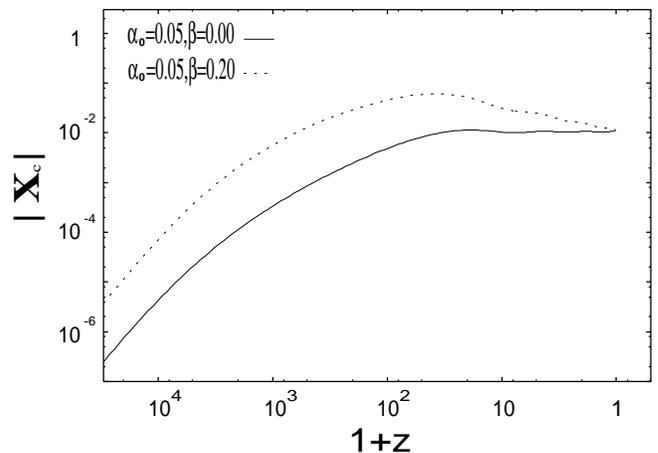}
\caption{The late time evolution of $X_c$ on a scale 
$k=0.003$ Mpc${}^{-1}$ in the scalar-tensor models 
with SCDM parameters. Normalization is arbitrary. 
These modes enter the horizon around $z \simeq 10^2$. 
Their amplitude is proportional to $\omega(\phi)^{-1}$ and 
the convergence at the present represents the same $\omega_0^{-1}$.} 
\label{fig:xxl}
\end{figure}

Fig.\ref{fig:xxs} and Fig.\ref{fig:xxl} show the early and the late time
behavior of the scalar field perturbation, respectively. 
The analytic estimation is presented in Appendix. 
Before the horizon entry, the scalar field fluctuation cumulates around
the dense region and $X_c$ always grows. Then, the amplitude is 
approximately proportional to the factor $\omega(\phi)^{-1}$ 
which regulates the nonminimal interaction. While, after the horizon 
entry, the field fluctuation flows out and its growth is suppressed. 
Then, $X_c$ oscillates at the same frequency as radiation fluid in the
radiation-dominated epoch or converges to some amplitude in the 
matter-dominated epoch. 

\begin{figure}
\includegraphics[width=\hsize]{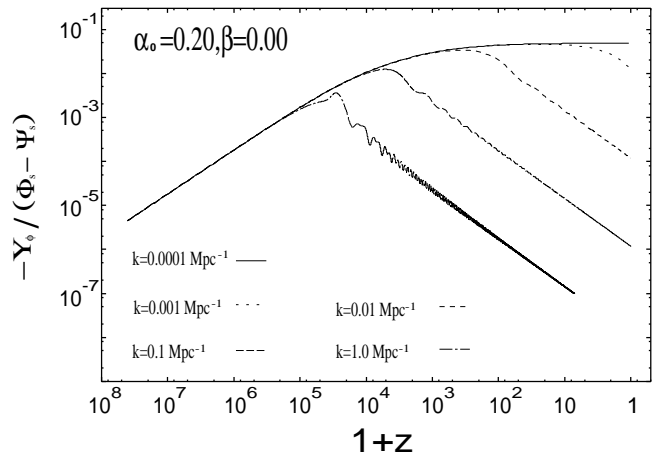}
\caption{The evolution of $ - \Upsilon_{\phi} / (\Phi_s-\Psi_s)$ 
on some typical scales in the Brans-Dicke model with SCDM parameters. 
Small scale modes are not affected by the scalar fluctuation.}
\label{fig:bdups_p}
\end{figure}

\begin{figure}
\includegraphics[width=\hsize]{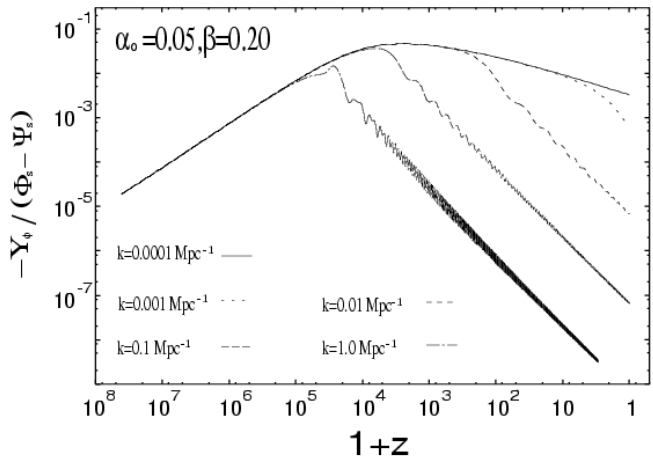}
\caption{The evolution of $ - \Upsilon_{\phi} / (\Phi_s-\Psi_s)$ 
on some typical scales in the attractor model with SCDM parameters. 
There exists the scale which is most affected by the scalar fluctuation.}
\label{fig:ups_p}
\end{figure}

As in the background, the contribution of the scalar field to the metric
is initially negligible. Until the mode enters the horizon, 
it becomes larger and reaches  $O({\omega_{ini}}^{-1})$ fraction
around the equality time. It is shown in Fig.\ref{fig:bdups_p} and 
Fig.\ref{fig:ups_p} where the feature of the largest scale mode is 
quite similar to the respective curve in Fig.\ref{fig:hub_evo}. 
Immediately after the horizon entry, the contribution suddenly drops 
and becomes negligible. This potential drop at the horizon entry is the 
characteristic event caused by the scalar field fluctuation 
especially in the matter-dominated epoch during which the horizon entry
does not involve the potential decay in the Einstein model. 
Because of the late time convergence to Einstein gravity, large scale
modes, as well as small scale modes which enter the horizon long before 
the equality time, are not affected by the scalar fluctuation in the
attractor model. 


\subsubsection{Metric perturbation}
\begin{figure}
\includegraphics[width=\hsize]{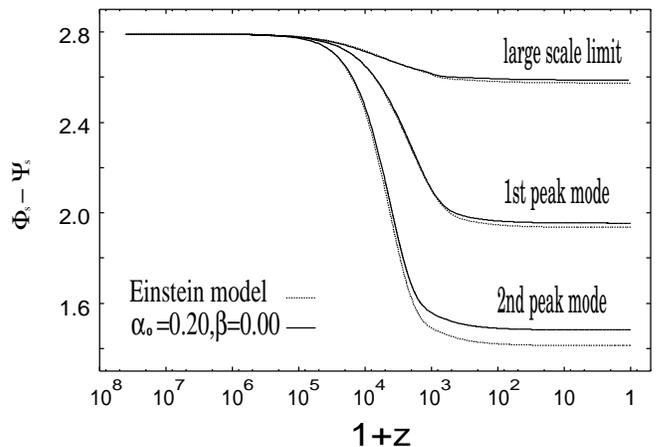}
\caption{The evolution of $\Phi_s-\Psi_s$ on some scales 
in the Einstein and the Brans-Dicke model with SCDM parameters. 
Normalization is arbitrary.}
\label{fig:psi_bd}
\end{figure}

\begin{figure}
\includegraphics[width=\hsize]{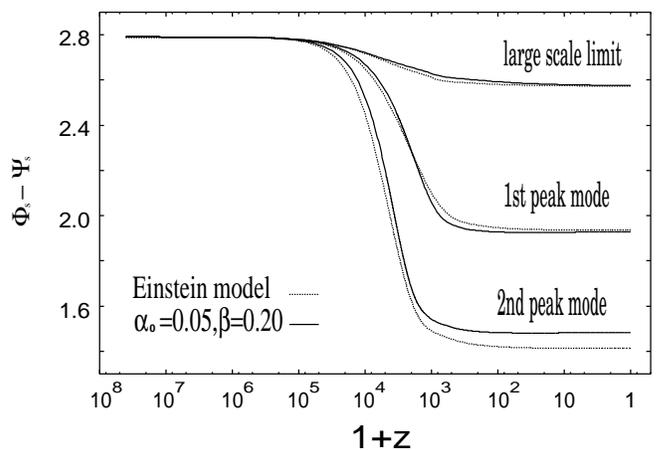}
\caption{The evolution of $\Phi_s-\Psi_s$ on some scales 
in the Einstein and the attractor model with SCDM parameters. 
Normalization is arbitrary.}
\label{fig:psi_st}
\end{figure}

\begin{figure}
\includegraphics[width=\hsize]{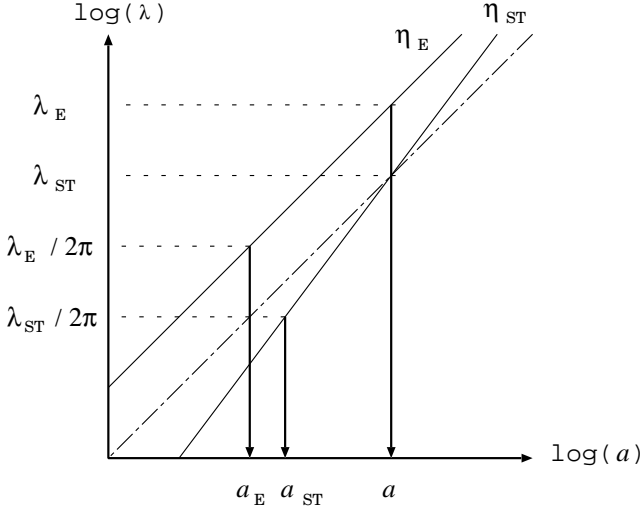}
\caption{The schematic diagram illustrating how the variation of $\phi$
  reduces the potential decay width, where $ a_0 / a = 1+z$. 
For instance, let $\lambda_E$ and 
$\lambda_{ST}$ be the second peak scales at some redshift in the Einstein 
and the scalar-tensor model, respectively. 
Due to the smaller $\phi$ in the past, the horizon expansion in the
scalar-tensor model behaves as lower solid line in the figure rather 
than dot-dashed line and hence the horizon entry is shifted to lower $z$
than that in the Einstein model.}
\label{fig:ills}
\end{figure}

The metric evolution directly affects the evolution of matter 
perturbations. Its evolution can be seen through the observations of the
present matter structures. For example, the turnover scale in the matter 
power spectrum corresponds to the horizon scale at the equality time. 
CMB observations provide more detailed information about the gravity 
evolution. Therefore, we pick up some scale modes which appear on the
CMB anisotropy spectrum as the characteristic structure, 
namely the modes corresponding to the acoustic peaks. 
The evolution of the first and the second acoustic peak modes at the
decoupling time is displayed in Fig.\ref{fig:psi_bd} and 
Fig.\ref{fig:psi_st} where the large scale limit modes are also compared
for reference. The initial behavior is model independent because the
scalar-tensor gravity behaves like Einstein gravity. 
In the Brans-Dicke model, the late time evolution, except for the
horizon entry, is flat as discussed in Appendix. 
Also in the attractor model, the late time evolution gradually becomes
flat because of the convergence to Einstein gravity. 
The large scale limit mode especially is not affected by the difference
in the horizon entry and its late time deviation is 
determined by $\omega(\phi)^{-1}$. After the equality, the decay 
widths of peak modes at a time generally become smaller in the
scalar-tensor models than those in the Einstein model (see e.g. the
second peak mode in Fig.\ref{fig:psi_bd} or Fig.\ref{fig:psi_st}). 
We point out two reasons for it. One reason is that the 
non-relativistic matter amount ($\Omega_0$) is larger in the
scalar-tensor models. The relative contribution of radiation to metric 
perturbations is smaller at the same $z$ and hence the decay caused 
by radiation pressure becomes smaller. The other reason is the motion 
of $\phi$. As explained in Fig.\ref{fig:ills}, if $\phi$ is constant, 
a peak mode in each model enters the horizon at the same time. 
However, since $\phi$ moves with time after the equality, 
the entry time of the peak mode in the scalar-tensor model is shifted 
to lower $z$. This time lag also makes the decay width smaller. 
Furthermore, during the early phase of potential decay (approximately
from the horizon entry up to the first maximum compression of photon
fluid), the contribution of the scalar induced decay occurring coherently 
to the decay deriving from radiation pressure is not negligible. 
For instance, the potential of the first peak mode in
Fig.\ref{fig:psi_bd} or Fig.\ref{fig:psi_st} catches up with that in 
the Einstein model around $z \sim 10^{3}$. 

\begin{figure}
\includegraphics[width=\hsize]{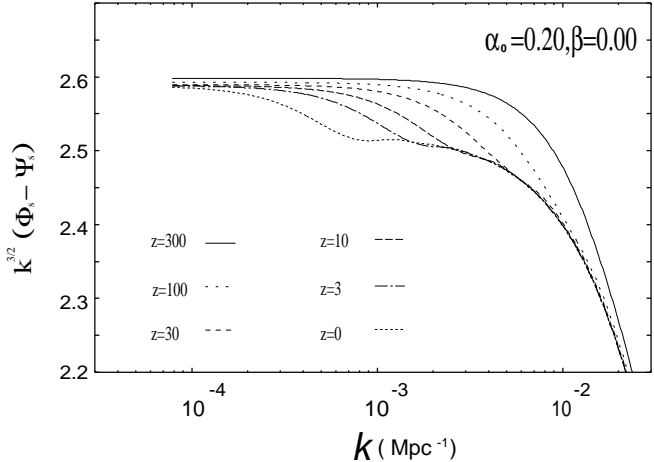}
\caption{The power spectra of gravitational potential $\Phi_s-\Psi_s$ 
at some redshifts in the Brans-Dicke model with SCDM parameters. 
Normalization is arbitrary. The potential decays to some extent as the
mode enters the horizon.}
\label{fig:bdup}
\end{figure}

\begin{figure}
\includegraphics[width=\hsize]{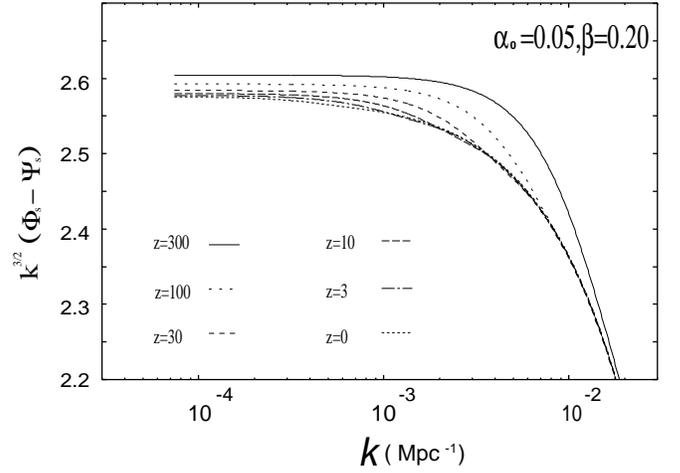}
\caption{The power spectra of gravitational potential $\Phi_s-\Psi_s$ 
at some redshifts in the attractor model with SCDM parameters. 
Normalization is arbitrary. The decay width becomes smaller as $\omega$
becomes larger.}
\label{fig:stup}
\end{figure}

The late time evolution of the power spectra of the gravitational
potential in the large scale in the scalar-tensor models is displayed
in Fig.\ref{fig:bdup} and Fig.\ref{fig:stup}. 
Since that scale enters the horizon much later after the equality
time, the scalar induced potential decay can be clearly seen. 
As noted above, the decay width is decreasing with time in the attractor
model and hence the large scale modes are not affected significantly. 


\subsubsection{Matter perturbation}
\begin{figure}
\includegraphics[width=\hsize]{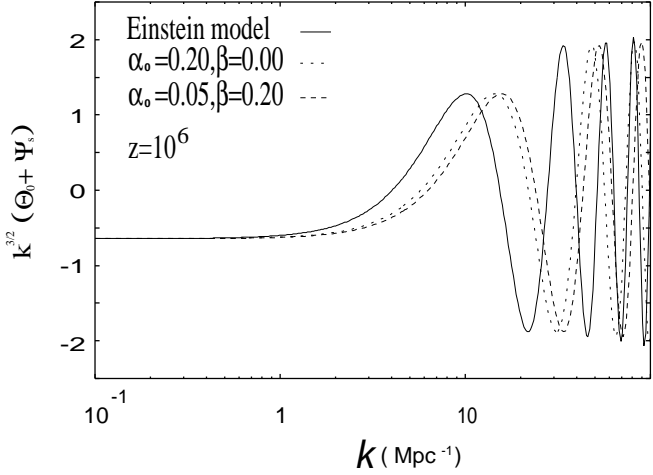}
\caption{The power spectra of effective temperature fluctuation at
$z=10^6$ in the Einstein and the scalar-tensor models with SCDM 
parameters. Normalization is arbitrary. Before $\phi$ begins to move, 
the only difference is that the gravitational "constant" 
in the scalar-tensor models is larger than Newton's constant. }
\label{fig:bdz106}
\end{figure}

In the radiation-dominated epoch, the difference in the magnitude of the
gravitational constant alone affects matter evolution. 
It is clearly seen in Fig.\ref{fig:bdz106} which exhibits the acoustic 
oscillation of photon fluid at $z=10^6$. As explained in 
Fig.\ref{fig:ills}, the peak mode in each model enters the horizon 
at the same time due to the constancy of the gravitational constant
($\phi$). Therefore, the degree of the peak boost is equal for each model. 
Parallel transports make these curves identical except in their
diffusion cut off. After the gravitational constant begins to vary, 
the peak heights generally become lower than those in the Einstein model. 
Exceptionally, during the first acoustic compression, the mode
experiences also the enhancement caused by the scalar induced potential
decay. 

\begin{figure}
\includegraphics[width=\hsize]{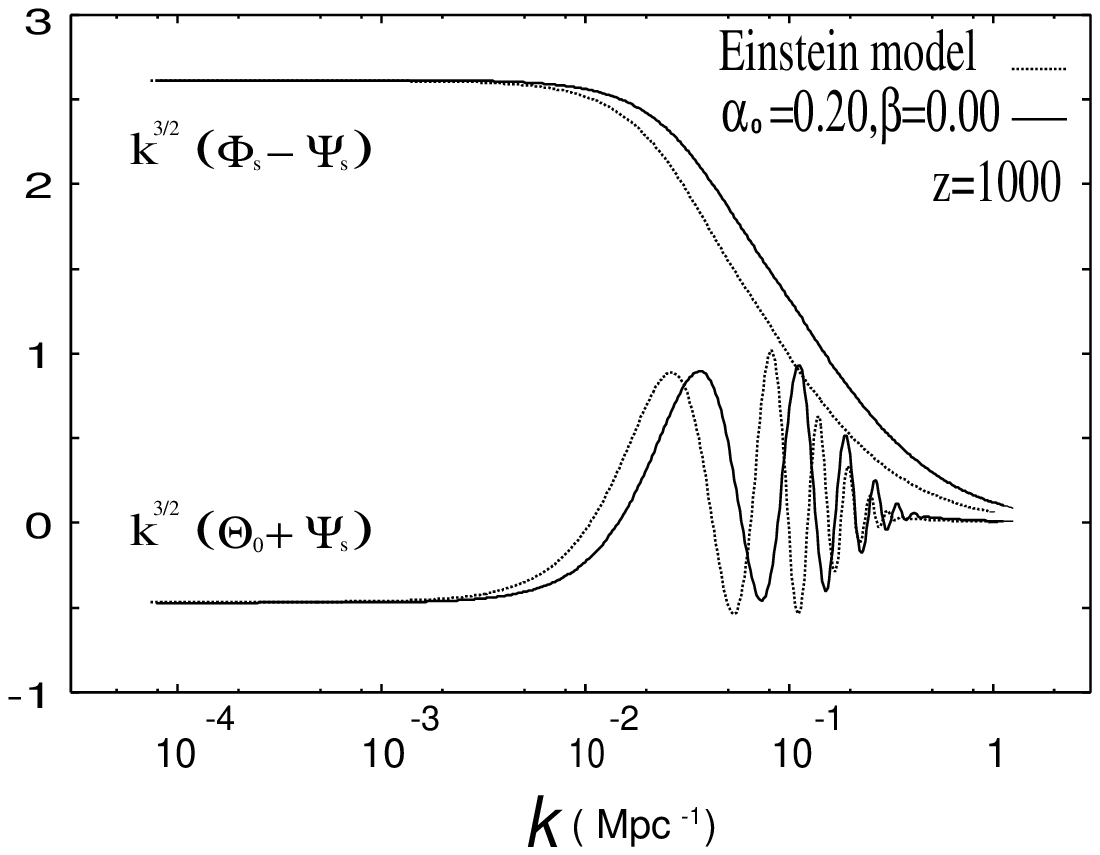}
\caption{The power spectra of effective temperature fluctuation 
$\Theta_0+\Psi_s$ and gravitational potential $\Phi_s-\Psi_s$ at
 $z=1000$ in the Einstein and the Brans-Dicke model with SCDM
 parameters. Normalization is arbitrary.}
\label{fig:bdz1000}
\end{figure}

\begin{figure}
\includegraphics[width=\hsize]{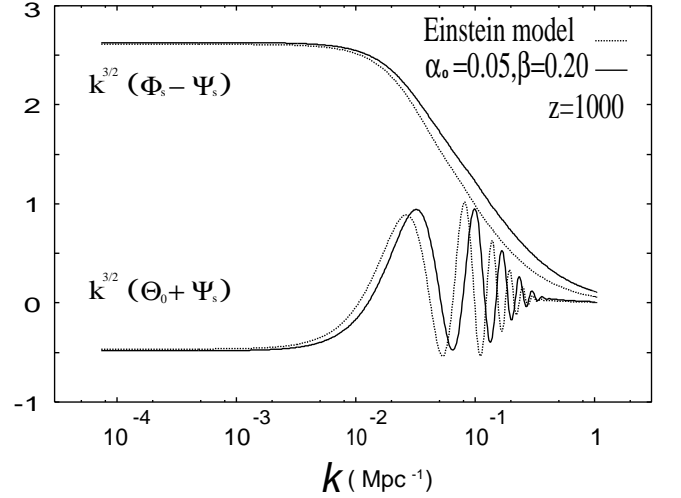}
\caption{The power spectra of effective temperature fluctuation 
$\Theta_0+\Psi_s$ and gravitational potential $\Phi_s-\Psi_s$ at
 $z=1000$ in the Einstein and the attractor model with SCDM
 parameters. Normalization is arbitrary.}
\label{fig:z1000}
\end{figure}

The observable oldest universe is the photon decoupling epoch. 
Fig.\ref{fig:bdz1000} and Fig.\ref{fig:z1000} show the photon and the
potential power spectra at $z=1000$. It should be noted that $z$ of 
the recombination time is not shifted because it is determined almost
only by the photon temperature. On very large scales (commonly called
Sachs-Wolfe plateau), the fluctuation amplitude in the scalar-tensor 
models is slightly larger than that in the Einstein model. The deviation 
represents the value of $\omega^{-1}$ at that time. The small scale 
potential height is, of course, higher in the scalar-tensor models, 
which implies the smaller horizon length at the equality time. 
On acoustic peak scales, in addition to peak locations which represent
the horizon length at the time, we can see three features: the first
Doppler peak height, the second peak height and the diffusion envelope. 
Let us begin with the diffusion scale. The diffusion length is also
dependent on the horizon length. In fact, it is defined as:
\begin{eqnarray}
l_D = \Bigl( \frac{l_{H}}{n_e \sigma_T}  \Bigr)^{\frac{1}{2}}
\Bigr|_{a=a_{rec}},
\end{eqnarray}
where $l_{H},n_e,$ and $\sigma_T$ are horizon length, free electron
density, and the Thomson scattering cross-section, respectively. And
$a_{rec}$ is the scale factor at the beginning of hydrogen recombination. 
The mean free path is model independent and hence 
the diffusion efficiency depends only on ${l_{H}}^{\frac{1}{2}}$. 
Therefore, the higher diffusion envelope in the scalar-tensor models also
derives from the smaller horizon length. 
On the other hand, single peak inside the envelope in the scalar-tensor 
models is lower than that in the Einstein model because the ratio of the peak
scales to the diffusion length is, in turn, proportional to
${l_{H}}^{-\frac{1}{2}}$. Aside from the diffusion effect, acoustic peak
heights in the scalar-tensor models generally become lower after the 
equality time due to the smaller potential decay of the peak modes. 
This results in the lower second peak in Fig.\ref{fig:bdz1000} and
Fig.\ref{fig:z1000}. This difference in the second peak height indeed 
originates from the difference in gravity, which can be confirmed by 
trying different $\Omega_b$ models. Concerning the first peak scale, 
as noted in the previous section, the scalar induced potential decay 
has non negligible effect. It occurs coherently to ordinary potential decay 
and makes the total decay width larger. This boosts the first 
Doppler peak higher because the mode is in its first maximum compression
phase at the decoupling time. (Precisely, ordinary potential decay is triggered by
a ``sound" horizon entry. Thus the degree of this enhancement correlates with 
the sound velocity of photon-baryon fluid. In larger $\Omega_b h^2$ models, 
the enhancement of the first peak becomes more significant.)

\begin{figure}
\includegraphics[width=\hsize]{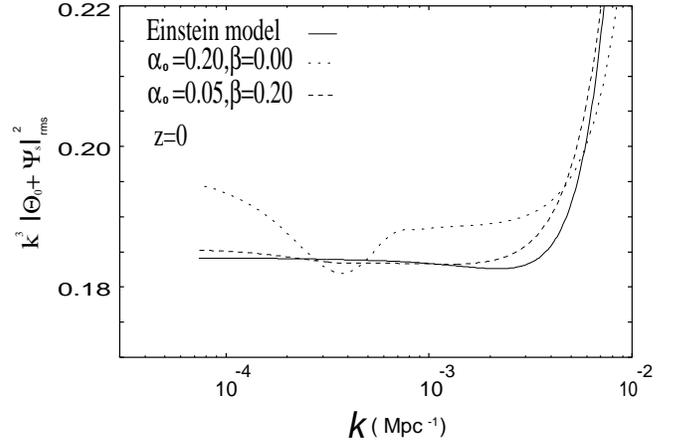}
\caption{The large scale power spectra of photon rms temperature 
fluctuation at the present in the Einstein and scalar-tensor models 
with SCDM parameters. Normalization is arbitrary. 
The scalar induced potential decay makes an uplift at the foot of the
first Doppler peak and a bump around the present horizon scale.}
\label{fig:rms}
\end{figure}

Also after the decoupling time, the scalar induced potential decay 
affects matter evolution at the horizon entry. It is found that 
it makes a bump in the potential transfer function (Fig.\ref{fig:bdup}
and Fig.\ref{fig:stup}) and the photon root mean square fluctuation
($|\Theta_0 + \Psi_s|^2_{rms} = |\Theta_0 + \Psi_s|^2 + \sum 
\frac{\Theta_l^2}{2l+1} $) spectrum (Fig.\ref{fig:rms}) around the
horizon scale of the time. 
In Fig.\ref{fig:rms}, the amplitude on super present horizon scales is 
higher than that in the Einstein model by $\omega_0^{-1}$. 
However, the amplitude deviation inside the horizon 
($k \sim 10^{-3}$ Mpc${}^{-1}$) is somewhat reduced by this ISW 
(integrated Sachs-Wolfe) effect. On the other hand, the temperature 
fluctuation at the foot of the first Doppler peak which is blue shifted 
at the decoupling time is enhanced by the effect. This uplift is not so
clear on the figure due to the smaller sound horizon length at the
decoupling. In the attractor model, this effect is not so large on
Sachs-Wolfe scales as on the scales around the first 
Doppler peak because the decay width is proportional to
$\omega(\phi)^{-1}$ at the horizon entry. 


\section{OBSERVATIONAL QUANTITIES}
Observations of the large scale structure of the Universe and 
the CMB anisotropies are the strong tools to constrain the theories of
primordial fluctuation evolution. Adopting scale invariant initial spectra, 
we present the matter power spectra and the CMB temperature anisotropy 
spectra based on the numerical calculations in the previous section. 

Before proceeding to the resulting spectra, we comment on the normalization. 
We find that, if the initial curvature perturbations are normalized to
the same amplitude, the temperature anisotropy power on COBE 
normalization angular scales is almost model independent. 
However, this does not necessarily imply that the large scale power on
the matter power spectrum is also model independent. 
The conservation of the adiabatic relation ($\Delta_{cr} / \Delta_{cm} = 4/3$) 
during when the mode is before the horizon entry is model independent. 
Here, the subscript $r$ and $m$ means a radiation and a
non-relativistic component, respectively. According to Appendix, 
this fact indicates that, in the Brans-Dicke model, the familiar 
Sachs-Wolfe relation is modified as follows:
\begin{eqnarray}
\Theta_0+\Psi_s = \frac{1}{3} \Bigl( 1 + \frac{2}{3\omega} \Bigr) \Psi_s
+ O(\omega^{-2}).
\end{eqnarray}
This is just an example. The actual modification also includes the
contribution of the ISW effect and others. 

\subsection{Matter power spectrum}
\begin{figure}
\includegraphics[width=\hsize]{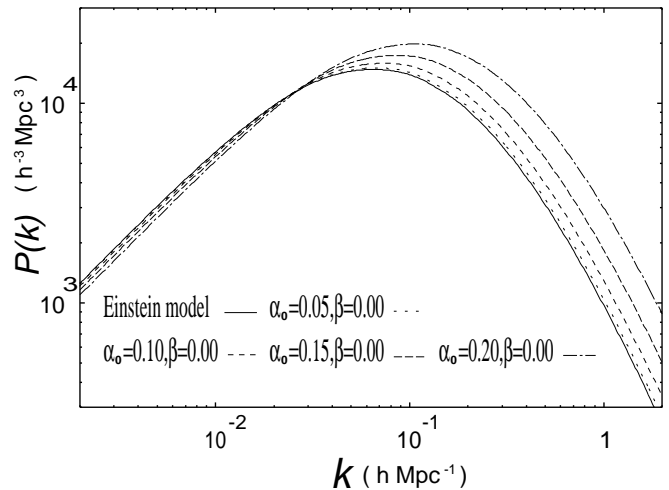}
\caption{The COBE normalized matter power spectra in the Brans-Dicke
  models with SCDM parameters. The turnover scale shifts 
according to the horizon scale at the matter radiation equality time. 
The difference in $\omega$ also affects the large scale amplitude.}
\label{fig:amps}
\end{figure}

\begin{figure}
\includegraphics[width=\hsize]{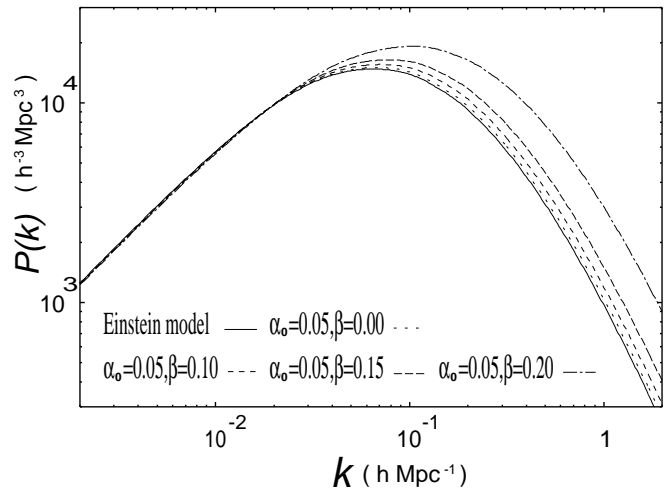}
\caption{The COBE normalized matter power spectra in the attractor
  models with SCDM parameters. The turnover scale shifts 
according to the horizon scale at the matter radiation equality time. 
The large scale amplitude is almost the same.}
\label{fig:mps}
\end{figure}

On subhorizon scales in the radiation-dominated epoch, 
the growth of density perturbations is inhibited by radiation pressure. 
After the transition from radiation domination to matter domination, 
perturbations on all length scales can grow by gravity. 
Hence in the matter perturbation spectrum, 
there exists the imprint of the horizon scale at the matter-radiation 
equality time. We can see the difference in the horizon scales 
at the equality time in Fig.\ref{fig:alp_evo} and Fig.\ref{fig:phi_evo}. 
The matter power spectra in the Einstein model and the scalar-tensor 
models are shown in Fig.\ref{fig:amps} and Fig.\ref{fig:mps}. 
It has been pointed out (in the context of Brans-Dicke cosmology
\cite{LMB}) that the variation of $\phi$ results in the larger expansion
rate and then the turnover scale of the present-day matter power spectrum
 shifts to the smaller scale, accordingly there is more small scale power. 
It is found from the figures that large scale power deviates from its
  Einstein counterpart in proportion to $\omega_0^{-1}$. 


\subsection{CMB temperature anisotropy spectrum}
\begin{figure}
\includegraphics[width=\hsize]{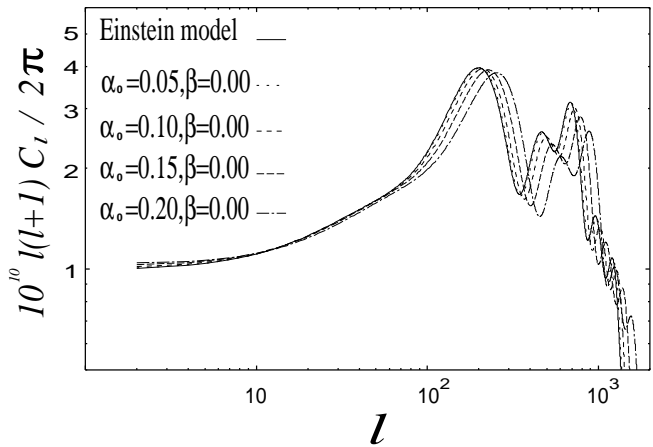}
\caption{The COBE normalized CMB temperature anisotropy spectra in the
  Brans-Dicke models with SCDM parameters.}
\label{fig:clas}
\end{figure}

\begin{figure}
\includegraphics[width=\hsize]{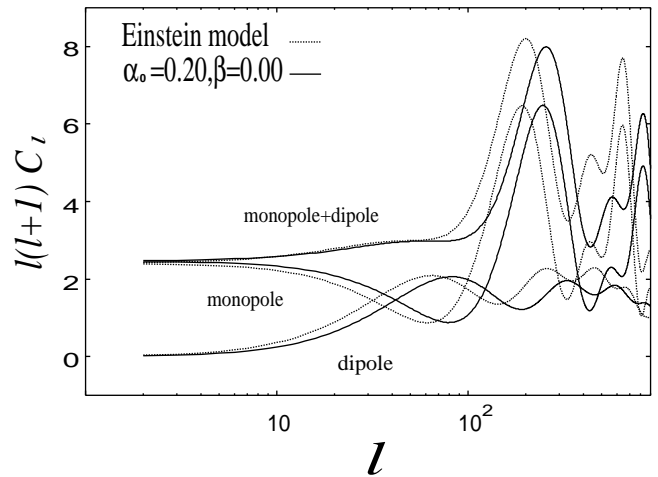}
\caption{The component decomposition of the CMB temperature anisotropy
  spectra in the Einstein and the Brans-Dicke model which is calculated 
by directly projecting the inhomogeneity spectra at the decoupling time 
into $l$ space. Normalization is arbitrary. Their curvature 
perturbations are initially normalized to be the same.}
\label{fig:sepa1}
\end{figure}

\begin{figure}
\includegraphics[width=\hsize]{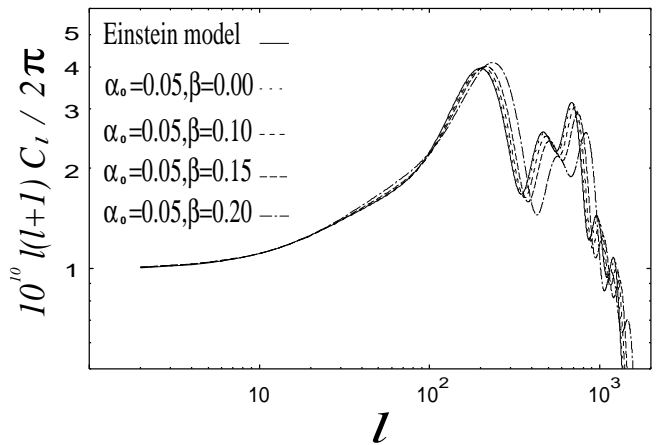}
\caption{The COBE normalized CMB temperature anisotropy spectra in the
  attractor models with SCDM parameters.}
\label{fig:cls}
\end{figure}

\begin{figure}
\includegraphics[width=\hsize]{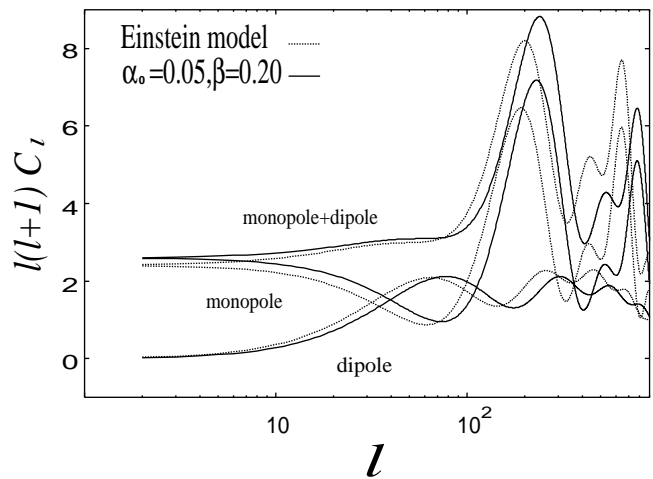}
\caption{The component decomposition of the CMB temperature anisotropy
  spectra in the Einstein and the attractor model which is calculated by
  directly projecting the inhomogeneity spectra at the decoupling time
  into $l$ space. Normalization is arbitrary. Their curvature 
perturbations are initially normalized to be the same.}
\label{fig:sepa2}
\end{figure}

The COBE normalized CMB anisotropy spectra are compared in
Fig.\ref{fig:clas} and Fig.\ref{fig:cls}. Fig.\ref{fig:sepa1} 
and Fig.\ref{fig:sepa2} show their component decomposition which is 
calculated by directly projecting the inhomogeneity spectra at the
decoupling time into $l$ space. 

First, we mention the shift of characteristic angular scales. 
The acoustic peak location (sound horizon scale) and the diffusion cut
off scale are dependent on the horizon length at the decoupling time. 
Since the matching condition of $\phi_0$ constricts the deviation of the
present horizon length, these angular scales directly represent the
horizon length at the decoupling. Therefore, the shift of peak locations 
to smaller angular scales is due to the large gravitational constant at
the decoupling. Although the damping scale also depends on the horizon
length, it is proportional to ${l_H}^{\frac{1}{2}}$. Hence the shift of
damping tail location is not so large as that of peak locations. It
means also that the width between the sound horizon scale and the diffusion
cut off scale becomes thinner. 

\begin{figure}
\includegraphics[width=\hsize]{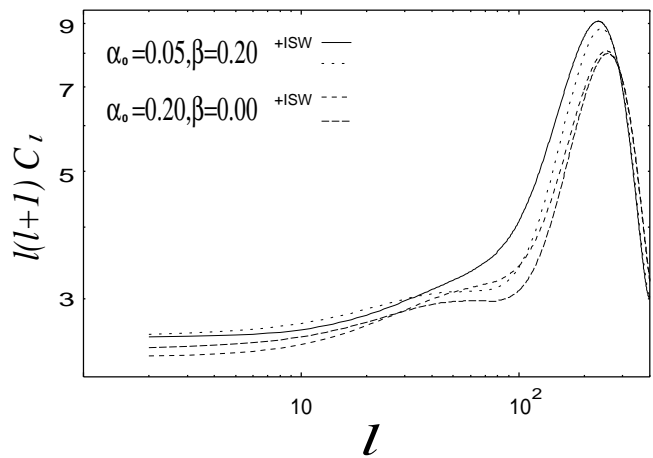}
\caption{The contribution of the scalar induced ISW effect to the CMB
  anisotropy spectra. The spectra not including the ISW contribution 
are calculated in the same way as in Fig.\ref{fig:sepa1} and 
Fig.\ref{fig:sepa2} and then the ISW contribution is added to them. 
The early ISW effect which is caused by radiation pressure is not
included in any curve.}
\label{fig:sisw}
\end{figure}

Next, let us proceed to the variation of fluctuation amplitude which
arises from several effects. On the largest angular scale, 
the amplitude deviates by $\omega_0^{-1}$ as in the matter power spectrum. 
The deviation on a Sachs-Wolfe scale is determined by the value of
$\omega(\phi)^{-1}$ at its horizon entry. In fact, the large scale 
tail in Fig.\ref{fig:cls} deviates not so large as that in
Fig.\ref{fig:sepa2} which represents the value of $\omega(\phi)^{-1}$ 
just at the decoupling time. Moreover, the scalar induced ISW effect 
is not negligible on large scales. We can see in Fig.\ref{fig:sisw} 
that it damps the fluctuation amplitude on the scales which are 
red shifted at the decoupling time. The smaller sound horizon scale at
the decoupling time also causes the attenuation of large scale 
amplitude in $l$ space. As just previously noted, the amplitude on 
COBE normalization scales is almost model independent. 
Probably the contributions of several effects cancel each other 
and the amplitude deviation around these scales is constricted. 
The foot before the first Doppler peak swells relatively to that in 
the Einstein model. The scalar induced decay contributes to this 
enhancement uplifting the originally blue shifted fluctuation, 
which can also be seen in Fig.\ref{fig:sisw}. This decay boosts the
acoustic peaks and the first peak is enhanced. 
Due to the peak location shift, each small scale peak enters the
diffusion envelope more deeply than the corresponding peak in the 
Einstein model and thus it is damped more effectively. 
As an aside, comparing Fig. 2 and Fig. 23, we expect that the difference
between the gravitational constant at the decoupling $G(z=1000)$ 
and at the present time $G_0$ may be constrained as 
$|G(z=1000)-G_0|/G_0 < 0.2$. The detailed discussion of observational
constraints of the model parameters is our future work.

In the precedent study devoted to the CMB anisotropy in Brans-Dicke
cosmology \cite{CK}, the Brans-Dicke models have higher acoustic 
peaks. On the contrary, according to our analysis, the Brans-Dicke models 
have generally lower acoustic peaks. In Fig.\ref{fig:sepa1} although 
the first peak height of the monopole component is almost the same as
that in the Einstein model, the attenuation of the dipole component 
on that angular scale reduces the total peak amplitude. 


\section{CONCLUSION}
We have comprehensively studied the perturbation evolution in the
scalar-tensor cosmological model. It is shown that the scalar-tensor 
gravity requires some modification to the standard evolution 
that is mainly caused by the variable gravitational constant and also
brings additional process derived from its fluctuation. 
We provide the insight into their influence on matter evolution and thus
it becomes possible to interpret the variation in the observable quantities. 

In the radiation-dominated epoch, the scalar-tensor gravity behaves like
Einstein gravity. The influence on the perturbation spectrum is the
larger horizon length alone. Its observational consequence is the
smaller turnover scale in the matter power spectrum. 
On the other hand, in the matter-dominated epoch, the nature of gravity
is qualitatively different from Einstein gravity. Consequently, 
aside from the shift of the acoustic peak locations and the diffusion
cut off scale, the enhancement and the attenuation which originate from 
the scalar field perturbation and others appear in the CMB anisotropy
spectrum. The attracting scalar-tensor gravity especially affects only
the small scale power because of its attracting nature 
and the degree is far beyond the post-Newtonian deviation at the present
epoch. 

In this paper, we have concentrated on studying the influence 
of the scalar-tensor gravity to cosmological perturbations and their
resultant observable power spectra. In the future work, we will put 
constraint on the model parameters $\alpha_0$ and $\beta$ by using the 
precise CMB and large scale structure data provided in near future. 

\begin{acknowledgments}
T.C. was supported in part by a Grant-in-Aid for Scientific 
Research (No.13740154) from the Japan Society for the Promotion of
Science and by a Grant-in-Aid for Scientific Research on Priority Areas 
(No.14047212) from the Ministry of Education, Science, Sports and 
Culture, Japan. 
N.S. is supported by the Alexander von Humboldt Foundation and 
a Japanese Grant-in-Aid for Science Research 
Fund of the Ministry of Education, No. 14540290. 
\end{acknowledgments}

\appendix*
\section{Analytic Approach}
In this Appendix, the analytic approximate solutions of both the
 background  and the perturbation equations in Brans-Dicke cosmology 
are presented (hence $\omega$ is constant here). 
Although these solutions are applicable only to Brans-Dicke models, 
we can use them for interpreting our numerical results of the variable $\omega$ models. 
The background cosmology is a flat SCDM model. 
To derive the approximate solutions, we expand the solutions 
perturbatively by $\omega^{-1}$. To the 0th order in $\omega^{-1}$,
we regard $\phi$ as constant, $\phi=1$. (Therefore, $X_c$ is vanished 
to the 0th order in $\omega^{-1}$.) 
This means we fix the homogeneous solution of Eq.(\ref{eq12}) to the 
0th order in $\omega^{-1}$. Hence the 0th order equations and 
their solutions are the same as those in the Einstein model.
The purpose of this section is to derive the corrections linear order
 in $\omega^{-1}$. Higher order corrections are neglected here. 
(Hence this treatment is applicable only in large $\omega$ models)

\subsection{Background}
Several authors have studied cosmological solutions in Brans-Dicke 
cosmology \cite{NARI}-\cite{HW}. 
Here, we show the cosmological solution which is valid from the 
radiation-dominated epoch to the matter-dominated epoch. 
In the background cosmology, the linear order in $\omega^{-1}$ is assigned to
$\phi'$, because the 0th order of $\phi$ is constant and the 
force term of Eq.(\ref{eq12}) is the linear order in $\omega^{-1}$.
Since Eq.(\ref{eq12}) is the linear order in $\omega^{-1}$, we can
substitute 0th order solutions to the coefficient of each term and
obtain the linear order
correction of the constant 0th order solution of $\phi$.
In both the radiation-dominated epoch and the matter-dominated epoch, 
the homogeneous solution of Eq.(\ref{eq12}) consists of $\phi'=0$, which 
corresponds to another choice of $G$ in Einstein gravity,
and a decaying mode. 
More generally, we must give the boundary conditions for Hubble parameter and matter energy density
to fix the magnitude of the decaying mode. 
However, we assume that the decaying mode is already negligible at $z \sim 10^8$
(the initial time of our numerical calculations), otherwise it
significantly affects the thermal history in the earlier epochs. 
(If the homogeneous decaying mode is dominant in 
$\phi'$, the energy density of $\phi$ is proportional to $a^{-6}$. 
It is stiffer than that of radiation fluid.) 
So we employ, as the main contribution 
to the solution of Eq.(\ref{eq12}), its particular solution that is
\begin{eqnarray}
\phi'=\frac{\sqrt{2}}{\omega} \frac{H_{eq} a_{eq}^3}{a^2}
\Biggl(1- \biggl( 1+\frac{a}{a_{eq}} \biggr)^{\frac{1}{2}}
\biggl(1-\frac{a}{2a_{eq}} \biggr) \Biggr), \label{eq63}
\end{eqnarray}
where $a_{eq}$ and $H_{eq}$ are the scale factor and Hubble parameter 
without $O(\omega^{-1})$ corrections at the matter-radiation equality time.
Specifically, this choice determines the deviation of $\Omega_0$ from unity. 
Integrating the above equation, we obtain
\begin{eqnarray}
\phi=\phi_0 + f(a)-f(a_0), \label{eq64}
\end{eqnarray}
where
\begin{eqnarray}
f(a)=\frac{2}{\omega}
\Biggl[ \frac{a_{eq}}{a}
-\frac{a_{eq}}{a} \biggl( 1+\frac{a}{a_{eq}} \biggr)^{\frac{1}{2}} \hspace{6.0em} \nonumber \\
+\frac{1}{2} \ln   \biggl\{ \frac{a}{a_{eq}} \cdot
\frac{ (1+\frac{a}{a_{eq}})^{\frac{1}{2}}+1 }
{ (1+\frac{a}{a_{eq}})^{\frac{1}{2}}-1 } \biggr\}
\Biggr], \label{eq65}
\end{eqnarray}
and $a_0$ is the present scale factor. 
The boundary value $\phi_0$ is $1+\frac{1}{2\omega}+O(\omega^{-2})$.

\subsubsection{Radiation-dominated epoch}
In the radiation-dominated epoch, Eq.(\ref{eq65}) is reduced to
\begin{eqnarray}
f(a) = \frac{1}{\omega}  \Biggl( \ln4 - 1 + \frac{3}{4} 
\frac{a}{a_{eq}} \Biggr). \label{eq66}
\end{eqnarray}
We neglect $O(a^2/a_{eq}^2)$ corrections here. 
In the period that $a/a_{eq} \ll 1$ is valid (i.e. early times in 
the radiation-dominated epoch), the last term 
can be neglected. This behavior of $\phi$ is well known 
\cite{NARI}. The contributions of $\phi'$ and time variation of $\phi$ 
to $a'/a$ are still smaller order than that of CDM energy density. 
This is because $\phi$ is driven via the nonminimal coupling 
whose amplitude is regulated by the energy density of non relativistic matter. 
Then, we obtain the same relation between $\eta$ and the 
expansion rate as that in Einstein gravity,
\begin{eqnarray}
\eta = {\Bigl( \frac{a'}{a}  \Bigr)}^{-1}.
\end{eqnarray}

\subsubsection{Matter-dominated epoch}
The solution in the matter-dominated epoch is,
\begin{eqnarray}
f(a) = \frac{1}{\omega} \ln (a/a_{eq}).
\end{eqnarray}
So Eq.(\ref{eq64}) becomes
\begin{eqnarray}
\phi=\phi_0 + \frac{1}{\omega} \ln (a/a_0). \label{eq69}
\end{eqnarray}
The widely known power-law solution
\begin{eqnarray}
\phi=\phi_0 \Bigl( \frac{a}{a_0} \Bigr)^{\frac{1}{\omega+1}} 
\end{eqnarray}
is reduced to Eq.(\ref{eq69}) with $O(\omega^{-2})$ corrections. 
Using the Eq.(\ref{eq69}), we obtain the following relation,
\begin{eqnarray}
\eta = 2 \Bigl(1-\frac{1}{\omega} \Bigr) {\Bigl( \frac{a'}{a}  \Bigr)}^{-1}.
\end{eqnarray}

\subsection{Perturbation}
In this section, introducing a new parameter
\begin{eqnarray}
x \equiv k \eta,
\end{eqnarray}
we derive the solutions of the perturbation equations. 
Obviously $x$ is smaller than unity on superhorizon scales 
and larger on subhorizon scales. 
$C_r$ and $C_m$ in the following sections are proportionality constants. 

\subsubsection{Radiation-dominated epoch}
We consider only the adiabatic initial conditions 
for matter perturbations in which $\Gamma = 0$.
Here we also drop the anisotropic pressure correction to simplify the problem. 
The 0th order initial conditions for matter perturbations 
are the same as those in Einstein gravity. 
Employing the parameter $x$ and neglecting the subdominant terms, 
Eq.(\ref{eq57}) is reduced to
\begin{eqnarray}
\frac{d^2}{dx^2} X_c + \frac{2}{x} \frac{d}{dx} X_c + X_c = 
 \frac{3}{8\omega x} \frac{a}{a_{eq}} \Bigl( \frac{d}{dx} \Delta_c \nonumber \\ 
 + \frac{1}{x} \Delta_c  \Bigr)
\label{eq72}
\end{eqnarray}
in the radiation-dominated epoch. The 0th order solution of $\Delta_c$ is
\begin{eqnarray}
\Delta^{(0)}_c / C_r = \Bigl( \frac{9\sqrt 3}{x} \sin (x / \sqrt 3 ) -
9\cos (x / \sqrt 3 )  \Bigr).
\label{eq73}
\end{eqnarray}
Then, as in the background case, we employ only the particular
solution of Eq.(\ref{eq72}) which is
\begin{eqnarray}
X_c / C_r = \frac{27 \sqrt 3}{16\omega x} \frac{a}{a_{eq}} \Biggl\{ 
\sin (x / \sqrt 3) \hspace{6em} \nonumber \\
 + \frac{\sqrt 3}{x} \Bigl( \cos x 
- \cos (x/ \sqrt 3) \Bigr) \Biggr\}.
\label{eq74}
\end{eqnarray}
(The particular solution is the only growing mode in $X_c$. 
If the homogeneous mode exists in $X_c$, it forms 
the gravitational potential perturbation decreasing with time.) 
The behavior of this solution is as follows
\begin{eqnarray}
X_c / C_r =\left\{ 
\begin{array}{ll}
\frac{3}{32\omega} (a/a_{eq}) x^2 & {\rm Superhorizon} \\
\frac{27}{16\omega} (a/a_{eq}) \frac{\sin (x / \sqrt 3)}{(x/\sqrt 3)}. &
{\rm Subhorizon} \\
\end{array}
\right.
\end{eqnarray}
Before the horizon entry, driven by the nonminimal coupling, $X_c$ grows
proportional to $x^2$. After the horizon entry, it stops growing and
begins to oscillate. The first term in the parenthesis in
Eq.(\ref{eq74}) $(\cos x)$ whose frequency is different from other terms
represents the propagation of $\phi$ fluctuation and damps as the
universe expands. The oscillation of the remaining terms is driven by 
the matter acoustic oscillation which affects $X_c$ via metric 
perturbations. The factor $1 / \sqrt{3}$ in their argument represents 
the radiation sound velocity. It is clear that the contribution of 
the additional terms in Eq.(\ref{eq56}) and (\ref{eq58}) to the 
gravitational potential is at most order $(a/a_{eq})\omega^{-1}$ which is 
still smaller order than that of CDM perturbation. Hence, at the 
initial time early enough, the initial conditions for matter 
perturbations are not modified. The relevant numerical solution is 
illustrated in Fig.\ref{fig:xxs}.

\subsubsection{Matter-dominated epoch}
In the matter-dominated epoch, the terms associated with $c_s^2$ or $w$
are neglected, and then 
Eq.(\ref{eq57}) is reduced to
\begin{eqnarray}
\frac{d^2}{dx^2} X_c + \frac{4}{x} \frac{d}{dx} X_c + X_c =
\frac{2}{\omega x} \frac{d}{dx} \Delta_c + \frac{6}{\omega x^2}
\Delta_c. \label{eq77}
\end{eqnarray}
The modified evolution equation for $\Delta_c$ becomes
\begin{eqnarray}
\frac{d^2}{dx^2} \Delta_c + \frac{2}{x} \frac{d}{dx} \Delta_c
-\frac{6}{x^2} \Delta_c = 
\frac{2}{\omega x}  \frac{d}{dx} \Delta_c \hspace{2em} \nonumber \\ 
- \frac{4}{\omega x^2} \Delta_c 
+ \frac{d^2}{dx^2} X_c + \frac{2}{x}
\frac{d}{dx} X_c -  \frac{6}{x^2} X_c. \label{eq78}
\end{eqnarray}
Here we use Eq.(\ref{eq77}). 
Obviously the homogeneous growing mode solution 
(the 0th order in $\omega^{-1}$) of Eq.(\ref{eq78}) 
is proportional to $x^2$. Let $C_m$ be the proportionality constant of it. 
Then, the particular solution of Eq.(\ref{eq77}) is 
\begin{eqnarray}
X_c / C_m = \frac{10 }{\omega} \Bigl( 1 - \frac{3 \sin x}{x^3} + \frac{3
  \cos x}{x^2}  \Bigr). 
\label{eq79}
\end{eqnarray}
The asymptotic forms are
\begin{eqnarray}
X_c / C_m =\left\{ 
\begin{array}{ll}
\frac{1}{\omega} x^2 & {\rm Superhorizon} \\
\frac{10 }{\omega}. & {\rm Subhorizon} \\
\end{array}
\right.
\end{eqnarray}
Before the horizon entry, it grows as in the radiation-dominated epoch. 
After the horizon entry, the propagation mode damps away and the
constant mode supported by the nonminimal coupling remains. 
The constancy of the surviving term is due to the constancy of the
gravitational potential. This convergence of $X_c$ is illustrated in 
Fig.\ref{fig:xxl}. Substituting Eq.(\ref{eq79}) into Eq.(\ref{eq78}), 
we obtain the modified solution of $\Delta_c$ as follows
\begin{eqnarray}
\Delta_c / C_m = x^2 + \frac{1}{\omega}  \Biggl\{ 
30 \Bigl( \frac{\cos x}{x^2} 
- \frac{\sin x}{x^3} \Bigr) \hspace{2em} \nonumber \\
+ 10 - x^2 \Biggr\}. \label{eq80}
\end{eqnarray}
The asymptotic forms of the correction terms are
\begin{eqnarray}
\Delta^{(1)}_c / C_m =\left\{ 
\begin{array}{ll}
-\frac{1}{840\omega} x^4 & {\rm Superhorizon} \\
-\frac{1}{\omega} x^2. & {\rm Subhorizon} \\
\end{array}
\right.
\end{eqnarray}
This correction is negligible until the horizon entry, and after the
horizon entry the amplitude of the density perturbation is slightly
suppressed. Then, the solution of the gravitational potential with
$O(\omega^{-1})$ accuracy becomes
\begin{eqnarray}
(\Phi_s - \Psi_s)/C_m = 12 + \frac{1}{\omega} \Biggl\{ 
60 \Bigl( \frac{3\sin x}{x^5} - \frac{3\cos x}{x^4} \nonumber \\ - \frac{\sin x}{x^3} \Bigr) - 26 \Biggr\}.
\end{eqnarray}
At the horizon crossing, transient decay is induced by the suppression
of $X_c$ growth. However, before and after the crossing time, the linear
order correction of the gravitational potential is flat as in the case
of the 0th order. 
\begin{eqnarray}
(\Phi_s - \Psi_s)^{(1)} / C_m =\left\{ 
\begin{array}{ll}
-\frac{22}{\omega} & {\rm Superhorizon} \\
-\frac{26}{\omega}. & {\rm Subhorizon} \\
\end{array}
\right.
\end{eqnarray}
Finally, we mention the proportionality constants. Comparing the same
scale modes in the two models ever since the radiation-dominated epoch, 
we obtain the different $C_m$s even if initially we set the $C_r$s to
the same value. $C_m$ itself contains the $O(\omega^{-1})$ correction
which represents the deviation formed around the matter-radiation
equality time. On superhorizon scales, the correction coefficient of the
proportionality constant is 
\begin{eqnarray}
C_m^{BD} / C_m^E \simeq 1+\frac{19}{10\omega}
\end{eqnarray}
according to our numerical calculations.

\end{document}